\documentclass[twocolumn,amsmath,nofootinbib]{revtex4-1}

\usepackage{url}
\usepackage{amssymb}
\usepackage{amsfonts}
\usepackage{amsbsy}
\usepackage{graphicx}
\usepackage{epstopdf}
\usepackage{hyperref}
\usepackage{array} 
\usepackage{multirow}

\newcolumntype{x}[1]{%
>{\centering\hspace{0pt}}p{#1}}%
\providecommand{\openone}{\leavevmode\hbox{\small1\kern-3.8pt\normalsize1}}

\begin{document}
\input{epsf.sty}


\title{Testing Multifield Inflation: A Geometric Approach}

\def\harvard{1}
\def\mit{2}
\def\affilmrk#1{$^{#1}$}

\author{
Courtney M. Peterson\affilmrk{\harvard},
Max Tegmark\affilmrk{\mit}
}

\address{
\parshape 1 -3cm 24cm
\affilmrk{\harvard} Dept.~of Physics, Harvard University, Cambridge, MA 02138, USA \\
\affilmrk{\mit} Dept.~of Physics \& MIT Kavli Institute, Massachusetts Institute of Technology, Cambridge, MA 02139
}

\date{Submitted to Journal Month day year, revised Month day, accepted Month day}
\date{November 21, 2012}

\vspace{10mm}

\begin{abstract}
We develop an approach for linking the power spectra, bispectrum, and trispectrum to the geometric and kinematical features of multifield inflationary Lagrangians.  Our geometric approach can also be useful in determining when a complicated multifield model can be well approximated by a model with one, two, or a handful of fields.  To arrive at these results, we focus on the mode interactions in the kinematical basis, starting with the case of no sourcing and showing that there is a series of mode conservation laws analogous to the conservation law for the adiabatic mode in single-field inflation.  We then treat the special case of a quadratic potential with canonical kinetic terms, showing that it produces a series of mode sourcing relations identical in form to that for the adiabatic mode.  We build on this result to show that the mode sourcing relations for general multifield inflation are extension of this special case but contain higher-order covariant derivatives of the potential and corrections from the field metric.  In parallel, we show how these interactions depend on the geometry of the inflationary Lagrangian and on the kinematics of the associated field trajectory.   Finally, we consider how the mode interactions and effective number of fields active during inflation are reflected in the spectra and introduce a multifield consistency relation, as well as a multifield observable $\beta_2$ that can potentially distinguish two-field scenarios from scenarios involving three or more effective fields. 
\end{abstract} 

\maketitle

\section{Introduction}

Inflation solves cosmic conundrums such as the horizon, flatness, and relic problems \cite{Guth-1981,Linde-1990,LythAndRiotto-1998,LiddleAndLyth-2000,BassettEtAl-2005}.  It also offers a mechanism for producing the primordial density fluctuations.  According to the inflationary paradigm, our Universe experienced an early period of quasi-exponential expansion that stretched quantum fluctuations beyond the causal horizon.  Once beyond the horizon, the fluctuations became locked in as classical perturbations, eventually initiating the formation of galaxies and large-scale structure \cite{MukhanovAndChibisov-1981,MukhanovAndChibisov-1982,Hawking-1982,Starobinksy-1982,GuthAndPi-1982,BardeenEtAl-1983}. 

Generically, inflation predicts that these classical perturbations should produce a small, nearly scale-invariant spectrum of primordial density fluctuations.  Measurements of the Cosmic Microwave Background (CMB), large-scale structure, supernovae, and gravitational lensing so far support the inflationary paradigm.  These measurements reveal that not only were the primordial fluctuations nearly scale-invariant, small, and include superhorizon fluctuations, but also that our Universe is essentially flat, as predicted by inflation (see \cite{KomatsuEtAl-2010} and references therein).  

But the ultimate goal is to use cosmic data not just to test the inflationary paradigm but to find the particular inflationary model that describes our Universe.  Of the myriad inflationary models that might describe our Universe, there is good reason to consider models where inflation is driven by multiple scalar fields.  First, many theories beyond the Standard Model---such as grand unification, supersymmetry, and effective supergravity from string theory---predict the existence of multiple scalar fields, which makes the presence of multiple fields likely during the hot, early Universe.  Second, multifield models have become increasingly popular in recent years.

\begin{figure}[h]
\vskip 22 pt
\centerline{\includegraphics[width=115mm]{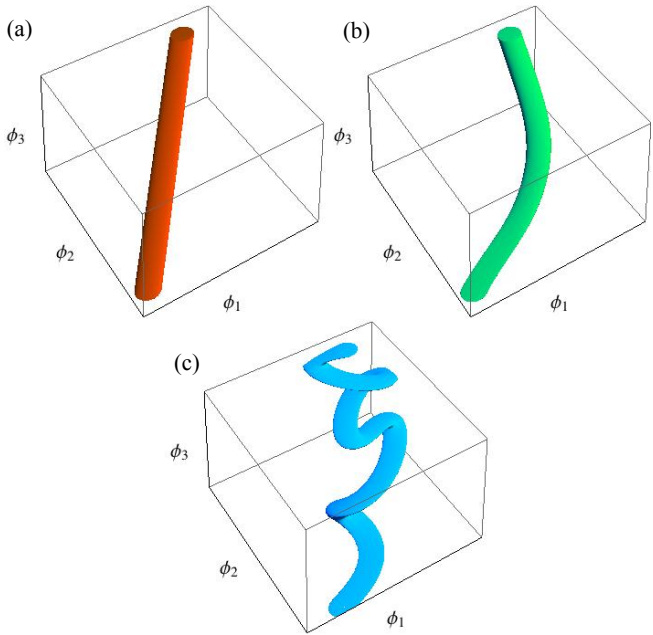}}
\vskip-3mm
\caption{Examples of three-field inflation trajectories that can be accurately approximated by an effective model with (a) one, (b) two, and (c) three fields, respectively.  For example, the field trajectory in (b) requires more than one effective field to represent the trajectory because it curves, but the trajectory curves in a plane, so only two effective fields are needed.}
\label{TubesFig}
\end{figure}

But the sobering reality of searching for multifield models that could describe our early Universe is that there is a staggeringly large number of multifield scenarios, making it impractical to test every scenario against cosmic data.  Unlike for single-field models, both the initial conditions and one or more Lagrangian parameters must be varied in order to fully test the range of scenarios arising from a given form of the Lagrangian.  We illustrated this point in \cite{PetersonAndTegmark-2010} by examining both two-field quadratic and power-law product potentials.  For each class of potentials, we tested more than 10,000 scenarios by varying both a parameter value in the Lagrangian and the initial conditions, in order to constrain the model using WMAP data on the power spectra.   Rigorously constraining models like this is extremely time-consuming.  

Rather than testing inflationary scenarios one by one like this, a more promising approach is to determine how constraints on the spectral observables in turn constrain the features of the inflationary Lagrangian.   Clearly, features such as the geometry of the inflationary potential influence the evolution of the field perturbations, so the spectra should constrain the geometry of the potential.  But in what ways do the spectra constrain the geometry of the inflationary potential?  Is there a way to tell from cosmic data whether a one-field or two-field model can fit all measurements, as illustrated in Figure~1, or whether more fields are required?     And what is the role of nonstandard kinetic terms in determining the cosmic observables?  In this paper, we aim to give greater insight into these and related questions. 

As background, initial work on understanding the perturbations and power spectra in general multifield inflation was done in \cite{KodamaAndSasaki-1984,SalopekEtAl-1989,Salopek-1995,SasakiAndStewart-1995,NakamuraAndStewart-1996,SasakiAndTanaka-1998,HwangAndNoh-2000,NibbelinkAndVanTent-2000,HwangAndNoh-2001,NibbelinkAndVanTent-2001,GongAndStewart-2002,VanTent-2003,RigopoulosEtAl-2005c,LeeEtAl-2005,LangloisAndRenaux-Petel-2008}.  The specific case of two-field inflation was treated in \cite{LangloisAndRenaux-Petel-2008,MukhanovAndSteinhardt-1997,GordonEtAl-2000,BartoloEtAl-2001b,BartoloEtAl-2001a,WandsEtAl-2002,DiMarcoEtAl-2002,DiMarcoAndFinelli-2005,ByrnesAndWands-2006,LangloisAndVernizzi-2006,LalakEtAl-2007,Renaux-PetelAndTasinato-2008,Gao-2009,PetersonAndTegmark-2010}.    Work towards calculating other spectra, such as the bispectrum and trispectrum, in general two-field and multifield inflation was done by \cite{RigopoulosEtAl-2004,RigopoulosEtAl-2005,RigopoulosEtAl-2005b,KimAndLiddle-2006,BattefeldAndEasther-2006,BattefeldAndBattefeld-2007,YokoyamaEtAl-2007,YokoyamaEtAl-2007b,MisraAndShukla-2008,Huang-2009a,ByrnesAndChoi-2010,TanakaEtAl-2010,KimEtAl-2010,BartoloEtAl-2001,BernardeauAndUzan-2002,BernardeauAndUzan-2002b,VernizziAndWands-2006,ChoiEtAl-2007,ByrnesEtAl-2008,VincentAndCline-2008,Wang-2010,MeyersAndSivanandam-2010,EllistonEtAl-2011,TzavaraAndTent-2010,Watanabe-2011}, among others. While developing formulas for the spectra from multifield inflation has received much attention, the sourcing relations among the modes in general multifield inflation where the Lagrangian is unspecified has received less attention.  The powerful $\delta N$ formalism introduced in \cite{SasakiAndStewart-1995} enables one to calculate the spectra in terms of gradients of the number of e-folds, $N$, but it has its limitations:  it can be applied analytically only to a fraction of models, and it does not provide any insight into the sourcing relations among modes.   
This situation contrasts with the case of general two-field inflation \cite{MukhanovAndSteinhardt-1997,GordonEtAl-2000,BartoloEtAl-2001b,BartoloEtAl-2001a,WandsEtAl-2002,DiMarcoEtAl-2002,DiMarcoAndFinelli-2005,ByrnesAndWands-2006,LangloisAndVernizzi-2006,LalakEtAl-2007,Renaux-PetelAndTasinato-2008,Gao-2009,PetersonAndTegmark-2010,LangloisAndRenaux-Petel-2008} and certain classes of multifield potentials (e.g., product potentials, sum potentials), where the mode interactions have been studied in depth. 

In this paper, we fill this important gap in the literature by examining the series of mode sourcing relations and how they reflect the geometric and kinematical properties of the inflationary Lagrangian.  This paper extends and complements some of our earlier work on two-field inflation \cite{PetersonAndTegmark-2010,PetersonAndTegmark-2010b}.    The rest of this paper is organized as follows.  In Sections \ref{Background EoM}-\ref{kinematics}, we cover the dynamics and kinematics of the background fields, and we discuss underappreciated subtleties of the slow-roll limit as it applies to multifield inflation in Section \ref{limits}.  In Sections \ref{perb EoMs}-\ref{mode equations}, we present equations of motion for the field perturbations in both the given and kinematical bases.  We then discuss mode evolution in the absence of sourcing and present mode conservation laws in Sections \ref{no sourcing}-\ref{mode cons laws}.  Section \ref{quad potls} treats the special case of quadratic potentials with canonical kinetic terms in which the mode sourcing equations radically simplify, and we use this as a reference point in Section \ref{sourcing} for deciphering how the mode sourcing relations in general multifield inflation depend on the geometric and kinematical features of the inflationary Lagrangian.  Finally, in Sections \ref{dimension}-\ref{nonGauss}, we use these sourcing equations to examine the effective number of fields in multifield models and to explore how this number is reflected in spectral observables.  We also generalize our two-field semianalytic formulas for the bispectrum and trispectrum \cite{PetersonAndTegmark-2010b} to multifield inflation, identify a spectral observable that can be used to distinguish two-field models from models with three or more fields (Section \ref{isocurv}), and introduce a new multifield consistency condition (Section \ref{nonGauss}).    This work helps pave the way towards a better understanding of how the cosmic observables can be used to constrain the form of the multifield inflationary Lagrangian.

\section{Background Fields}
\label{Background Fields}

This section covers the dynamics and kinematics of the background inflationary fields.  In turn, we review notation and the equation of motion for the background fields in Section \ref{Background EoM}, outline a framework for parsing the field vector kinematics in Section \ref{kinematics}, and cover the slow-roll and slow-turn limits in Section \ref{limits}.

\subsection{Background Field Equation}
\label{Background EoM}

We consider general multifield inflationary scenarios with the following characteristics.  Inflation is driven by an arbitrary number of scalar fields, $\phi_i$, where $i=1,2,...,d$, and $d$ is the total number of scalar fields present during inflation, not all of which may be contributing to the inflationary expansion at a given time.   We use Latin indices to represent quantities related to the fields, $\phi_i$, and we represent the fields compactly as
\begin{align}
\label{phi vector}
\boldsymbol{\phi} \equiv (\phi_1,\phi_2,...,\phi_d),
\end{align}
calling $\boldsymbol{\phi}$ the field vector for short, even though the fields do not transform as vectors.  During and after inflation, we assume Einstein gravity and that the non-gravitational part of the inflationary action is described by
\begin{align}
\label{action}
S = \int \left[-\frac{1}{2} g^{\mu \nu} G_{ij}(\boldsymbol{\phi})\ \frac{\partial \phi^i}{\partial x^{\mu}} \frac{\partial \phi^j}{\partial x^{\nu}} - V(\boldsymbol{\phi})\right] \sqrt{-g} \, d^4x,
\end{align}
where $g_{\mu \nu}$ is the spacetime metric, the fields are expressed in units of the reduced Planck mass, $\bar{m} \equiv 1/\sqrt{8\pi G}$, and $c = \hbar = \bar{m} = 1$.   The tensor $G_{ij}$ is a function of only the fields, and it determines the form of the kinetic terms in the Lagrangian;  it can be viewed as inducing a field manifold and hence is called the field metric.  If the kinetic terms are canonical, then $G_{ij} = \delta_{ij}$.  In this manuscript, we treat the case of \textit{general} multifield inflation, meaning we do not assume a particular functional form for either the field metric or the inflationary potential.  

Before proceeding, we introduce some notational shorthand.    For vectorial quantities lying in the tangent and cotangent bundles of the field manifold, we use boldface vector notation and standard inner product notation:
\begin{align}
\label{vector prod defn}
\mathbf{A}^{\dag} \mathbf{B} \equiv \mathbf{A} \cdot \mathbf{B} \equiv G_{ij} A^i B^j,
\end{align}
where we use the symbol $^{\dag}$ on a naturally contravariant or covariant vector to denote its dual, e.g., $\dot{\boldsymbol{\phi}}^{\dag} \equiv (G_{ij}\dot{\phi}^j)$ and $\boldsymbol{\nabla}^{\dag} \equiv (G^{ij} \nabla_j)$.  Also, instead of working in terms of the coordinate time, $t$, we work in terms of $N$, which represents the logarithmic growth of the scale factor, $a(t)$:
\begin{align}
\label{N defn}
dN \equiv d \ln a = H \, dt,
\end{align}
where $H \equiv \frac{\dot{a}}{a}$ is the Hubble parameter.  $N$ represents the number of e-folds of the scale factor, $a(t)$.  We work in terms of $N$ because it is dimensionless, it relates to a more physical measure of time, and it simplifies the equations of motion \cite{SasakiAndTanaka-1998, PetersonAndTegmark-2010}.  Differentiation with respect to $N$ is denoted by
\begin{align}
\label{d/dN shorthand}
' \equiv \frac{d}{dN}.
\end{align}  

The background equation of motion for the fields is derived by imposing covariant conservation of energy.   We derived such an equation using $N$ as the time variable for general two-field inflation in \cite{PetersonAndTegmark-2010}, and the same equation holds for the general case of multifield inflation:
\begin{align}
\label{Phi EoM wrt N}
\frac{\boldsymbol{\eta}}{(3 - \epsilon)} + \boldsymbol{\phi}' + \boldsymbol{\nabla}^\dag \ln V & = 0,
\end{align}
where
\begin{align}
\label{epsilon}
\epsilon \equiv - (\ln H)' =  \frac{1}{2} \boldsymbol{\phi}' \cdot \boldsymbol{\phi}',
\end{align}
and the covariant field acceleration $\boldsymbol{\eta}$ is defined as
\begin{align}
\label{eta defn}
\boldsymbol{\eta} \equiv \frac{D \boldsymbol{\phi}'}{dN}.
\end{align}
The symbol $D$ acting on a contravariant vector $X^i$ means
\begin{align}
\label{D}
DX^i \equiv d\phi^j \, \nabla_j X^i  = d\phi^j \left(\partial_j X^i + \Gamma^i_{jk} X^k \right), 
\end{align}
where $\Gamma^{i}_{jk}$ and $\nabla_j$ are the Levi-Civita connection and the covariant derivative, respectively, associated with the field metric.   Therefore, the covariant acceleration $\boldsymbol{\eta}$ represents deviations from perfect parallel transport of $\boldsymbol{\phi}'$.  By working in terms of $D$ and the covariant derivative $\boldsymbol{\nabla}$, we are able to write all the equations of motion in manifestly covariant form.

Finally, we make the common assumption that over the course of inflation, the field vector picks up speed but not necessarily monotonically.  Eventually, the field vector picks up enough speed to end inflation, which we take to be at $\epsilon = 1$. The choice of exactly when inflation ends does not impact the results presented in this manuscript.

\subsection{Field Vector Kinematics}
\label{kinematics}

The kinematical framework presented here is based mostly on work by  \cite{GordonEtAl-2000,NibbelinkAndVanTent-2000,NibbelinkAndVanTent-2001,PetersonAndTegmark-2010}, with small modifications.  Here, the coordinates are the scalar fields, which represent the coordinate position on the manifold induced by the field metric.  In analogy to Newtonian mechanics, $\boldsymbol{\phi}$ represents the position, $\boldsymbol{\phi}'$ is the velocity, and $\boldsymbol{\eta} \equiv \frac{D \boldsymbol{\phi}'}{dN}$ represents the covariant acceleration, where $\frac{D}{dN}$ is defined through Eq. (\ref{D}).   Similarly, we can define higher-order covariant derivatives of the field velocity.   The jerk is defined as
\begin{align}
\boldsymbol{\xi} \equiv \frac{D^2 \boldsymbol{\phi}'}{dN^2}.
\end{align}
An equation of motion for the jerk can be obtained by differentiating Eq. (\ref{Phi EoM wrt N}) once, which yields 
\begin{align}
\label{jerk EoM}
\frac{\boldsymbol{\xi}}{(3 - \epsilon)} + \boldsymbol{\eta} = - \left[\mathbf{M} + \frac{\boldsymbol{\eta} \boldsymbol{\eta}^\dag}{(3 - \epsilon)^2} \right] \boldsymbol{\phi}',
\end{align}
where the \textit{mass matrix}, $\boldsymbol{M}$, is defined as
\begin{align}
\label{M defn}
\mathbf{M} \equiv \boldsymbol{\nabla}^\dag \boldsymbol{\nabla}\ln V
\end{align}
and is symmetric.  Similarly, we represent the $(n-1)$-th covariant derivative of the velocity by the notation\footnote{For comparison, Nibbelink and Van Tent \cite{NibbelinkAndVanTent-2001} defined a series of higher-order kinematical vectors as
\begin{align}
\label{van Tent vecs}
\boldsymbol{\tilde{\eta}}^{(n)} \equiv \frac{\mathcal{D}^{(n-1)} \boldsymbol{\phi}^{;}}{\left(\frac{a^;}{a}\right)^{(n-1)} |\boldsymbol{\phi}^{;}|},
\end{align}
where $;$ represents the derivative with respect to the arbitrary time variable $\tau$ and $\mathcal{D}$ is the ``slow-roll derivative''.  The ``slow-roll derivative'' is defined as $\mathcal{D}(b^n A) \equiv (\frac{D}{d\tau} - n \frac{d \ln b}{d\tau}) (b^n A)$, where $b = - g_{00}$ and $A$ is independent of $b$.  Our kinematical vectors differ from Nibbelink and Van Tent's in two ways: (1) the effective order in the slow-roll expansion, which differs because of the factor of $|\boldsymbol{\phi}^{;}|$ in the denominator in Eq. (\ref{van Tent vecs}), and (2) the expressions themselves---that is, our series differs from theirs even when the order of Eq. (\ref{van Tent vecs}) is adjusted by multiplying by $|\boldsymbol{\phi}^{;}|$.  Both constructs have their utility: Nibbelink and Van Tent's construct makes their vectors manifestly independent of the choice of time coordinate, while our construct is physically intuitive since it is based on using $N$ as the time variable and can be used to simplify certain expressions to a greater degree.}  
\begin{align}
\label{kine vecs}
\boldsymbol{\chi}^{(n)} = \left(\frac{D}{dN}\right)^{(n-1)} \boldsymbol{\phi}',
\end{align}
and an equation of motion for $\boldsymbol{\chi}^{(n)}$ can be obtained by differentiating Eq. (\ref{Phi EoM wrt N}) a total of $n-2$ times.

These kinematical vectors induce a basis in which the perturbed equations of motion can be better understood \cite{GordonEtAl-2000,NibbelinkAndVanTent-2000,NibbelinkAndVanTent-2001}. The construction of this basis is as follows.  The first basis vector, $\mathbf{e}_1$, is chosen to lie in the direction of the field velocity, parallel to the field trajectory.  The second basis vector, $\mathbf{e}_2$, is constructed to lie along the part of the field acceleration that is orthogonal to the field velocity, in the direction that makes $\mathbf{e}_2 \cdot \boldsymbol{\eta} \ge 0$.  Continuing the Gram-Schimdt orthogonalization procedure produces a set of $d$ basis vectors:
\begin{align}
\label{basis vecs}
\mathbf{e}_1 \equiv & \frac{\boldsymbol{\phi}'}{|\boldsymbol{\phi}'|}, \nonumber \\
\mathbf{e}_2 \equiv & \frac{(\mathbf{I} - \mathbf{e}_1 \mathbf{e}_1^{\dag}) \boldsymbol{\eta}}{\vert(\mathbf{I} - \mathbf{e}_1 \mathbf{e}_1^{\dag}) \boldsymbol{\eta}\vert}, \nonumber \\
. . .  \\
\mathbf{e}_d \equiv & \frac{\left(\mathbf{I} - \sum\limits_{i=1}^{d-1} \mathbf{e}_i \mathbf{e}_i^{\dag}\right) \boldsymbol{\chi}^{(d)}}{\left\vert \left(\mathbf{I} - \sum\limits_{i=1}^{d-1} \mathbf{e}_i \mathbf{e}_i^{\dag}\right) \boldsymbol{\chi}^{(d)}\right\vert},  \nonumber 
\end{align}
where $\mathbf{I}$ is the identity matrix of the appropriate dimensionality.   If, however, one of the kinematical vectors $\boldsymbol{\chi}^{(n)}$ already lies in the subspace defined by the basis vectors $\mathbf{e}_1, \mathbf{e}_2,...,\mathbf{e}_{n-1}$, then it is not possible to find a projection of $\boldsymbol{\chi}^{(n)}$ that represents a new direction in field space.  In this case, $\mathbf{e}_n$ can simply be constructed at will so that it represents a new direction that is orthogonal to the subspace spanned by the basis vectors $\mathbf{e}_1$ through $\mathbf{e}_{n-1}$, and then the orthogonalization process can naturally proceed again.  While our kinematical vectors differ from those of Nibbelink and Van Tent \cite{NibbelinkAndVanTent-2000,NibbelinkAndVanTent-2001}, our kinematical basis vectors are equivalent to theirs.

\begin{figure}[pbt]
\centerline{\includegraphics[width=78mm]{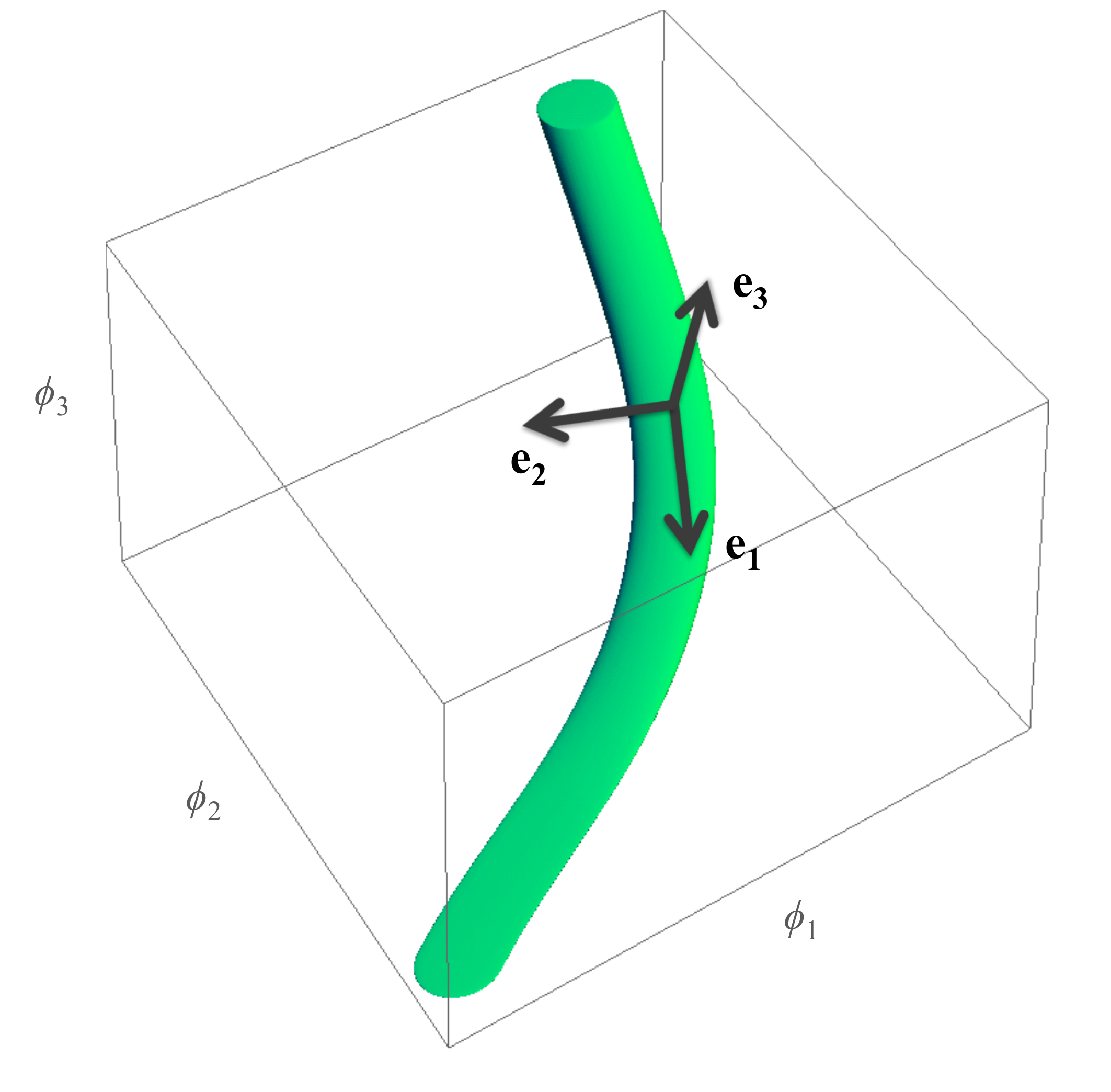}}
\vskip-3mm
\caption{An example showing how the kinematical basis is constructed from the kinematics of the background fields.  The green curved path represents the trajectory of the field vector in a three-field inflation scenario.  At each point on the trajectory, the $\mathbf{e}_1$ basis vector is chosen to point in the direction of the field velocity, $\mathbf{e}_2$ points in the direction of the perpendicular acceleration, and $\mathbf{e}_3$ is constructed to be orthogonal to the first two basis vectors.}
\label{KineBasisFig}
\end{figure}

With these basis vectors, we can take projections of vectors and matrices.  For example,
\begin{align}
\chi^{(m)}_n \equiv \mathbf{e}_n \cdot \boldsymbol{\chi}^{(m)}
\end{align}
represents the projection of the $m$th kinematical vector onto the $n$th basis vector.  Note that because of the definition of the kinematical basis vectors in Eq. (\ref{basis vecs}), if $n > m$, then $\chi^{(m)}_n = 0$.  That is, in the kinematical basis, $\boldsymbol{\phi}'$ has the sole nonzero component 
\begin{align}
\label{v defn}
v \equiv |\boldsymbol{\phi}'|;
\end{align} 
$\boldsymbol{\eta}$ has nonzero components $\eta_1$ and $\eta_2$; $\boldsymbol{\xi}$ has nonzero components $\xi_1$, $\xi_2$, and $\xi_3$; and so on.  The projection of any vector along $\mathbf{e}_1$ is particularly noteworthy, as it represents the vector component parallel to the field trajectory and hence single-field-like behavior. Whereas vector components orthogonal to the field trajectory relate to effects unique to multifield inflation.   For this reason, it is useful to use the shorthand notation
\begin{align}
A_{mn} \equiv \mathbf{e}_m^\dag \boldsymbol{A} \mathbf{e}_n
\end{align}
for the matrix coefficients of any matrix $\mathbf{A}$ and 
\begin{align}
\label{perp matrix projection}
\mathbf{A}_{\perp \perp}  \equiv \left(\mathbf{I} - \mathbf{e}_1 \mathbf{e}_1^\dag\right) \mathbf{A} \left(\mathbf{I} - \mathbf{e}_1 \mathbf{e}_1^\dag\right)
\end{align}
for the above special matrix projection.

Lastly, we consider the time derivatives of the kinematical basis vectors, which represent how quickly the basis vectors are covariantly changing direction with respect to the field manifold.  In particular, the derivative of the basis vector parallel to the field velocity, $\mathbf{e}_1$, represents how quickly the field trajectory itself is covariantly changing direction and is given by 
\begin{align}
\label{e_1 derivative}
\frac{D \mathbf{e}_1}{dN} = \frac{\eta_2}{v} \mathbf{e}_2.
\end{align}
Similarly, differentiating the second basis vector in Eq. (\ref{basis vecs}) gives
\begin{align}
\label{e_2 derivative}
\frac{D\mathbf{e}_2}{dN} = \frac{\xi_3}{\eta_2} \mathbf{e}_3 - \frac{\eta_2}{v} \mathbf{e}_1.
\end{align}
The derivative of the $n$th basis vector is
\begin{align}
\label{e_n derivative}
\frac{D \mathbf{e}_n}{dN} = \frac{\chi^{(n+1)}_{n+1}}{\chi^{(n)}_n} \mathbf{e}_{n+1} - \frac{\chi^{(n)}_{n}}{\chi^{(n-1)}_{n-1}} \mathbf{e}_{n-1}. 
\end{align}
In analogy to our work in \cite{PetersonAndTegmark-2010}, we call $|\frac{D\mathbf{e}_n}{dN}|$ the \textit{turn rate for the nth basis vector}.  Note that Eq. (\ref{e_n derivative}) means that when the $\mathbf{e}_n$ basis vector changes direction, it can pick up components along only the $\mathbf{e}_{n-1}$ and $\mathbf{e}_{n+1}$ directions.  We emphasize that this fact will greatly simplify the equations of motion for the field perturbations.   Furthermore, because $\frac{D}{dN} (\mathbf{e}_{n+1} \cdot \mathbf{e}_n) = 0$, the matrix 
\begin{align}
\label{Z matrix}
Z_{mn} \equiv \mathbf{e}_m \cdot \frac{D\mathbf{e}_n}{dN} 
\end{align}
is skew-symmetric with the only nonzero components being
\begin{align}
Z_{n+1,n} = -Z_{n,n+1} = \frac{\chi^{(n+1)}_{n+1}}{\chi^{(n)}_{n}}.
\end{align}
The kinematical quantity $Z_{n+1,n}$ represents how quickly the $\mathbf{e}_{n}$ basis vector is turning into the direction of $\mathbf{e}_{n+1}$.  Because $\mathbf{Z}$ summarizes the turn rates for all $d$ basis vectors, we call $\mathbf{Z}$ the \textit{turn rate matrix}.  The turn rate matrix is therefore the multifield generalization of the idea of a single covariant turn rate for two-field inflation.   The turn rate matrix, along with the kinematical basis vectors, plays a key role in determining the dynamics of the field perturbations.

\subsection{Slow-Roll And Slow-Turn Limits}
\label{limits}

The final element of the background solution is the ``slow-roll limit,'' the standard approximation invoked when the fields are slowly rolling and the inflationary expansion is quasi-exponential.   In this section, we uncover some important nuances and make some distinctions regarding different formulations of the slow-roll approximation that have been assumed to be equivalent.

In multifield inflation, the slow-roll limit is typically defined (e.g., \cite{MukhanovAndSteinhardt-1997,BartoloEtAl-2001a,WandsEtAl-2002,DiMarcoAndFinelli-2005,ByrnesAndWands-2006}) by the two conditions:
\begin{align}
\label{trad slow-roll cond v1}
\epsilon \ll 1, \\
\label{trad slow-roll cond}
\left\vert M_{ij} \right\vert \ll 1,
\end{align}
which forces the field vector to be slow-rolling and the masses of the fields and their couplings to be small.  In other approaches (e.g., \cite{SasakiAndTanaka-1998,NibbelinkAndVanTent-2000,NibbelinkAndVanTent-2001}), the second condition above is effectively replaced by  
\begin{align}
\label{SRA alt}
\boldsymbol{\eta} \ll \boldsymbol{\phi}',
\end{align}
which more narrowly forces the dimensionless field acceleration to be much smaller than the dimensionless field velocity.

In \cite{PetersonAndTegmark-2010}, we examined the above conditions in the case of two-field inflation and argued for a more nuanced approach that splits the second slow-roll condition in Eq. (\ref{SRA alt}) into two separate limits---the slow-roll limit and the slow-turn limit.\footnote{In two-field inflation, the conditions in Eq. (\ref{trad slow-roll cond}) differ from those in Eq. (\ref{SRA alt}) by the extra constraint $M_{22} \ll 1$.}  We defined the two-field slow-roll limit as
\begin{align}
\label{new slow-roll cond v1}
\epsilon \ll 1, \\
\label{new slow-roll cond}
\left\vert \frac{\eta_1}{v} \right\vert \ll 1, 
\end{align}
which is identical to the single-field definition;  that is, Eqs. (\ref{new slow-roll cond}) and (\ref{new slow-roll cond}) correspond to limits on single-field-like behavior.  We elevated the second component of Eq. (\ref{SRA alt}) into a separate limit dubbed the \textit{slow-turn limit}: 
\begin{align}
\label{two-field slow-turn limit}
\left\vert \frac{D \boldsymbol{e}_1}{dN} \right\vert = \frac{\eta_2}{v} = Z_{21} \ll 1.
\end{align}
It corresponds to limits on how quickly the field trajectory is covariantly changing direction---a distinctly multifield behavior.  The power of our distinction is that the rolling (single-field behavior) and turning (multifield behavior) of the field vector have very different effects on the power spectra.  For example, two-field models that strongly violate the slow-turn limit around horizon-crossing but not the slow-roll limit are ruled out by WMAP constraints on the density power spectrum \cite{PetersonAndTegmark-2010} and can potentially produce large isocurvature modes, depending on the amplitude of the entropy mode \cite{PetersonAndTegmark-2010b}.  

To extend this more nuanced approach to general multifield inflation, now multiple turn rates must be taken into consideration.  We say that a basis vector $\mathbf{e}_n$ is slowly turning if
\begin{align}
\label{e_n slow-turn limit}
\left\vert \frac{D \mathbf{e}_n}{dN} \right\vert \ll 1.
\end{align}
When all $d$ basis vectors are slowly turning, we say that the inflationary scenario is in the \textit{slow-turn limit}, and the magnitude of every component of the turn rate matrix, $\mathbf{Z}$, is significantly less than one:
 \begin{align}
\label{Z slow-turn limit}
\left\vert Z_{ij} \right\vert \ll 1.
\end{align}

We claim that both the conditions in Eqs. (\ref{trad slow-roll cond}) and (\ref{Z slow-turn limit}) are needed to correctly analogize the slow-roll limit from single-field inflation to general multifield inflation.  One might wonder why Eq. (\ref{Z slow-turn limit}) is needed in addition to Eq. (\ref{trad slow-roll cond}), since in two-field inflation, Eq. (\ref{trad slow-roll cond}) implies Eq. (\ref{Z slow-turn limit}).  The answer is that while $Z_{21} \approx -M_{21}$, in general it is not true that $|M_{ij}| \approx |Z_{ij}|$.   For this reason, a total of three limits (or four if the field metric is nontrivial) are needed to simplify the perturbed equation of motion in a manner similar to that in single-field inflation.  These three conditions are the slow-roll limit Eq. (\ref{trad slow-roll cond v1}), the small mass limit in Eq. (\ref{trad slow-roll cond}), and the slow-turn limit in Eq. (\ref{Z slow-turn limit}); an additional limit on the curvature of the field manifold for nontrivial field metrics will be introduced in the next section.   These subtle but more important points have not been fully recognized before, to our knowledge.  Nonetheless, in this paper, we will refer to these four limits as the multifield slow-roll approximation to avoid introducing new nomenclature that might create confusion.  

Having clarified the correct analogous slow-roll conditions for multifield inflation, we now apply these three limits to the background equations of motion and to the perturbed equations of motion in Section \ref{Perbs}.  Eq. (\ref{Phi EoM wrt N}) for the evolution of the fields reduces to
\begin{align}
\label{Phi EoM slow-roll}
\boldsymbol{\phi}' \approx - \boldsymbol{\nabla}^\dag \ln V,
\end{align}
and the field speed is given by
\begin{align}
v \approx \left\vert \boldsymbol{\nabla} \ln V \right\vert,
\end{align}
or equivalently by
\begin{align}
\epsilon \approx \frac{1}{2} \left\vert \boldsymbol{\nabla} \ln V \right\vert^2.
\end{align}
By virtue of Eq. (\ref{Phi EoM slow-roll}), the operator $\frac{D}{dN} = \boldsymbol{\phi}' \cdot \boldsymbol{\nabla}$ becomes
\begin{align}
\frac{D}{dN} \approx -\boldsymbol{\nabla} \ln V \cdot \boldsymbol{\nabla},
\end{align} 
and the kinematical vectors can be approximated by
\begin{align}
\label{chi in slow-roll}
\boldsymbol{\chi}^{(n)} \approx \left(-\boldsymbol{\nabla} \ln V \cdot \boldsymbol{\nabla} \right)^{(n-1)} (-\boldsymbol{\nabla}^\dag \ln V).
\end{align}
For example, 
\begin{align}
\label{eta slow-roll v1}
\boldsymbol{\eta} \approx - \mathbf{M} \boldsymbol{\phi}'  \approx \mathbf{M} \boldsymbol{\nabla}^\dag \ln V.
\end{align}
Note the special result that follows from Eq. (\ref{eta slow-roll v1}):
\begin{align}
\label{eta slow-roll}
\frac{\eta_2}{v} = Z_{21} \approx M_{12}.
\end{align}
The approximations for the basis vectors follow directly from the above results, with Eq. (\ref{chi in slow-roll}) substituted for $\boldsymbol{\chi}^{(n)}$ in Eq. (\ref{basis vecs}); in particular,
\begin{align}
\mathbf{e}_1&  \approx - \frac{\boldsymbol{\nabla}^\dag \ln V}{|\boldsymbol{\nabla} \ln V|}.
\end{align}
In later sections, we use $=$ instead of $\approx$ and simply indicate when the slow-roll limit applies.  These results will help simplify the equations of motion and the interactions among modes.

\section{Field Perturbations}
\label{Perbs}

In this section, we show how the evolution of the field perturbations is determined by the kinematics of the background fields and the geometries of the potential and field manifold.  While this has been done in detail for general two-field inflation and for subcases of multifield inflation such as product potentials, our goal here is to provide the first more thorough treatment in the \textit{general} multifield case.  In Sections \ref{perb EoMs} and \ref{mode equations}, we present an equation of motion for the field vector perturbation in terms of the time variable $N$ in both the given and kinematical bases, respectively.  In Sections \ref{no sourcing}-\ref{sourcing}, we uncover how the mode interactions are determined by the kinematics of the field trajectory and geometric features of the inflationary Lagrangian.  We start with the case of no sourcing in Section \ref{no sourcing} and develop a series  of mode conservation laws in Section \ref{mode cons laws}.  In Section \ref{quad potls}, we treat the special and very interesting case of quadratic potentials with canonical kinetic terms where all mode equations greatly simplify and assume the same form.  We use this case as a reference for exploring the mode interactions in general multifield inflation in Section \ref{sourcing}.

\subsection{Field Vector Perturbation Equation}
\label{perb EoMs}

Here we work exclusively in the flat gauge.  In this gauge, the field perturbations decouple from the metric perturbations and equal the gauge-invariant Mukhanov-Sasaki variable $\boldsymbol{\delta \phi}_f = \boldsymbol{\delta \phi} + \psi \boldsymbol{\phi}'$, where $\psi$ represents the scalar metric perturbation on spatial hypersurfaces \cite{Sasaki-1986,Mukhanov-1988}. From here forward, we drop the subscript $f$ from $\boldsymbol{\delta \phi}_f$ for simplicity. 

The equation for the field perturbations is obtained by imposing covariant conservation of energy, as has been demonstrated before \cite{SasakiAndStewart-1995,NakamuraAndStewart-1996, NibbelinkAndVanTent-2000}.  However, we break from convention by using $N$ as the time variable, both because it is physically intuitive and it makes the equation of motion dimensionless.  We derived such an equation in the context of general two-field inflation with noncanonical kinetic terms in \cite{PetersonAndTegmark-2010}.  Following the same series of steps as outlined in \cite{PetersonAndTegmark-2010}, we arrive at the same expression, with the exception of the curvature term arising from the field metric is more complicated, reflecting the additional field degrees of freedom.\footnote{For comparison, in two-field inflation, the curvature term effectively reduces to a single degree of freedom, the Ricci scalar, times a scaled outer product of two kinematical basis vectors.}   The resulting equation in Fourier space is
\begin{align}
\label{Delta Phi EoM wrt N}
\frac{1}{(3 - \epsilon)} \frac{D^2\boldsymbol{\delta \phi}}{dN^2} + \frac{D\boldsymbol{\delta \phi}}{dN} & +  \left(\frac{k^2}{a^2 V}\right) \boldsymbol{\delta \phi} \nonumber \\ & = - \left[\mathbf{\tilde{M}} + \frac{\boldsymbol{\eta}\boldsymbol{\eta}^\dag}{(3-\epsilon)^2}\right] \boldsymbol{\delta \phi},
\end{align}
where $k$ is the comoving wavenumber.   The term $\mathbf{\tilde{M}}$ is the \textit{effective mass matrix},\footnote{For comparison to our dimensionless definition of the effective mass matrix, Nibbelink and Van Tent defined the effective mass matrix as  $\tilde{\mathbf{M}}^2 \equiv \boldsymbol{\nabla} \boldsymbol{\nabla}^\dag V - H^2 \mathbf{R}$ \cite{NibbelinkAndVanTent-2001}.} and we define it as
\begin{align}
\label{M eff defn}
\mathbf{\tilde{M}} \equiv \mathbf{M} - \frac{1}{(3-\epsilon)} \mathbf{R},
\end{align}
and the curvature matrix, $\mathbf{R}$, is defined as \cite{NibbelinkAndVanTent-2000}  
\begin{align}
\label{R defn}
R^a_{\, \, d} \equiv 2\epsilon \, R^a_{\, \, bcd} e_1^b e_1^c,
\end{align}
where $R^a_{\, \, bcd}$ is the Riemann curvature tensor associated with the field metric.  Because of the symmetry and anti-symmetry properties of the Riemann curvature tensor, it follows that $\mathbf{R}$ is symmetric.  Moreover, $\mathbf{R} \, \boldsymbol{\phi}' = 0$, and hence $\mathbf{R} = \mathbf{R}_{\perp \perp}$. 

Now we simplify Eq. (\ref{Delta Phi EoM wrt N}) for use in the superhorizon limit.  In this limit, the modes are significantly outside the horizon such that $\left(\frac{k}{aH}\right)^2 \ll 1$ and the subhorizon term $\left(\frac{k^2}{a^2 V}\right) \boldsymbol{\delta \phi}$ can be neglected.  We also invoke the multifield slow-roll approximation, as correctly outlined in Section \ref{limits}.  To start, the term $\frac{\boldsymbol{\eta} \boldsymbol{\eta}^\dag}{(3-\epsilon)^2}$ can be neglected since this term is much smaller than $\mathbf{M}$ as long as the field trajectory is not turning rapidly.  In brief, the simplification follows from the fact that $\boldsymbol{\eta} \approx  - \mathbf{M} \boldsymbol{\phi}$ in the multifield slow-roll limit, as we showed in more depth in \cite{PetersonAndTegmark-2010} for two-field inflation. We also introduce another simplifying condition, which completes the extrapolation of the slow-roll limit to the multifield case: like for $\mathbf{M}$ and $\mathbf{Z}$, the dimensionless components of $\mathbf{R}$ must satisfy
\begin{align}
\label{R SRA cond}
|R_{ij}| \ll 1.
\end{align}
Whenever all these conditions apply in the superhorizon limit, it can be shown \cite{TaruyaAndNambu-1997,NibbelinkAndVanTent-2001} that the acceleration of the field perturbations can also be neglected and Eq. (\ref{Delta Phi EoM wrt N}) reduces to
\begin{align}
\label{Delta Phi EoM wrt N slow-roll}
\frac{D\boldsymbol{\delta \phi}}{dN} = - \mathbf{\tilde{M}} \, \boldsymbol{\delta \phi},
\end{align}
where 
\begin{align}
\label{Delta Phi EoM wrt N slow-roll}
\tilde{\mathbf{M}}= \mathbf{M} - \frac{1}{3} \mathbf{R}
\end{align}
in the slow-roll limit. Equation (\ref{Delta Phi EoM wrt N slow-roll}) shows that the superhorizon, slow-roll evolution of the modes is determined by only $\mathbf{M}$, which can be viewed as the covariant couplings of the fields, and $\mathbf{R}$, which encapulates the curvature of the field manifold.  We point out that the simplicity of this equation follows from working in terms of the time variable $N$ and hence justifies our choice to depart from convention.

\subsection{Mode Evolution Equations in the Kinematical Basis}
\label{mode equations}

Now we examine the interactions among the $d$ modes in the superhorizon limit.  These interactions have been studied in depth for general two-field inflation and for certain classes of multifield potentials (e.g., product potentials, sum potentials)  \cite{MukhanovAndSteinhardt-1997,GordonEtAl-2000,BartoloEtAl-2001b,BartoloEtAl-2001a,WandsEtAl-2002,DiMarcoEtAl-2002,DiMarcoAndFinelli-2005,ByrnesAndWands-2006,LangloisAndVernizzi-2006,LalakEtAl-2007,Renaux-PetelAndTasinato-2008,Gao-2009,PetersonAndTegmark-2010,TzavaraAndTent-2010,Watanabe-2011,MeyersAndSivanandam-2011}.  But surprisingly,  studying the interactions among modes one by one like this has not received much attention in the \textit{general} multifield case.  Here we fill this important gap in the literature. 

In general, the interactions among modes are most easily understood in the kinematical basis.  First, in this basis, the density mode can be teased out from the $d$ modes.  Second, it turns out that the turn rate matrix can simplify in this basis. In the kinematical basis, the $n$th mode is represented as
\begin{align}
\delta \phi_n \equiv \mathbf{e}_n \cdot \boldsymbol{\delta \phi};
\end{align}
that is, the modes are decomposed by their projections along the kinematical basis vectors.  The adiabatic or density mode corresponds to $\delta \phi_1$, the component of $\boldsymbol{\delta \phi}$ that is parallel to the field trajectory.  The remaining components of $\boldsymbol{\delta \phi}$ in the kinematical basis correspond to entropy modes, represented collectively as $\boldsymbol{\delta \phi}_{\perp}$.  Entropy modes are linear combinations of the field perturbations that leave the overall density unperturbed.  As there are $d$ fields in the system, there will be $d-1$ entropy modes, all of which are orthogonal to the field trajectory and to each other.   We note that when a mode $\delta \phi_{m}$ affects the evolution of mode $\delta \phi_n$, we say that $\delta \phi_m$ \textit{sources} $\delta \phi_n$, regardless of whether that interaction causes $\delta \phi_n$ to increase or decrease in amplitude.  Since we group all sourcing terms on the right-hand side of each equation, we refer to such equations as \textit{mode sourcing equations}.   

We start with the well-known mode sourcing equation for the adiabatic mode, $\delta \phi_1$, which is most easily derived from the fact that the comoving density perturbation vanishes in the superhorizon limit.  Imposing this constraint yields \cite{GordonEtAl-2000,NibbelinkAndVanTent-2001,PetersonAndTegmark-2010} 
\begin{align}
\label{delta phi_1 exact EoM}
\left(\frac{\delta \phi_1}{v}\right) ' = 2 \frac{D\mathbf{e}_1}{dN} \cdot \frac{\boldsymbol{\delta \phi}}{v}  = 2 Z_{21} \left(\frac{\delta \phi_2}{v}\right).
\end{align}
In the slow-roll limit, the above equation can be written as
\begin{align}
\label{delta phi_1 derivative slow-roll}
\delta \phi_1' + M_{11} \delta \phi_1 = 2 Z_{21} \delta \phi_2,
\end{align}
where we have used that $Z_{21} \approx - M_{12}$ in Eq. (\ref{eta slow-roll}). Examining Eq. (\ref{delta phi_1 exact EoM}) reveals that the adiabatic mode is sourced only when the field trajectory changes direction with respect to the field manifold  (e.g., \cite{LythAndRiotto-1998,GordonEtAl-2000,NibbelinkAndVanTent-2000,NibbelinkAndVanTent-2001,PetersonAndTegmark-2010}).  The strength of the sourcing depends on the $\mathbf{e}_1$ turn rate: the faster the background trajectory changes direction, the faster the adiabatic density mode grows.\footnote{In the classical treatment of Eq. (\ref{delta phi_1 exact EoM}), this statement is implied to be taken with respect to assuming that $\delta \phi_1$ and $\delta \phi_2$ are both positive.  In the quantum treatment, when we speak of the mode $\delta \phi_1$ growing, it is implied that we are referring to the variance of $\delta \phi_1$ growing.  Similar assumptions are made when discussing the other modes.}  Moreover, the adiabatic mode can be sourced only by the entropy mode $\delta \phi_2$; none of the other modes can source the adiabatic mode.  Otherwise, when the field trajectory does not turn or $\delta \phi_2$ vanishes, it follows that $\delta \phi_1 \propto v$, which is tantamount to single-field behavior.   

For each of the entropy modes, we can also derive a mode sourcing equation.   We start from Eq. (\ref{Delta Phi EoM wrt N slow-roll}), which is a vector equation but takes a different form in the kinematical basis because the basis vectors can rotate, causing the equation of motion (\ref{Delta Phi EoM wrt N}) to pick up extra terms that trivially vanish in the original given basis.  Using the fact that
\begin{align}
\delta \phi_n' =  \mathbf{e}_n \cdot \boldsymbol{\delta \phi}' - \mathbf{e}_n^\dag \mathbf{Z} \, \boldsymbol{\delta \phi},
\end{align}
it follows that the corresponding equation in the kinematical basis is 
\begin{align}
\label{Delta Phi EoM slow-roll kine basis}
\frac{D\boldsymbol{\delta \phi}}{dN} \approx - \left[\mathbf{\tilde{M}} + \mathbf{Z} \right] \boldsymbol{\delta \phi}.
\end{align}
This quite simple but elegant equation is equivalent to the corresponding equation derived in \cite{NibbelinkAndVanTent-2001}, but tells us more straightforwardly that the evolution of modes depends only on the turn rate matrix and the effective mass matrix, which includes corrections from any nontrivial field metric.   Now projecting out the adiabatic mode and using Eq. (\ref{eta slow-roll}), the evolution of the $d-1$ entropy modes is described by
\begin{align}
\label{perp perb EoM}
\frac{D\boldsymbol{\delta \phi}_{\perp}}{dN} \approx - \left[\mathbf{\tilde{M}} + \mathbf{Z}\right]_{\perp \perp} \, \boldsymbol{\delta \phi}_{\perp},
\end{align}
where the special matrix projection was defined in Eq. (\ref{perp matrix projection}).  Eq. (\ref{perp perb EoM}) shows that the adiabatic mode does not source any of the entropy modes in the superhorizon limit---a result that holds true regardless of whether the slow-roll limit applies.  

The individual equations for each of the $d-1$ entropy modes form a series of mode sourcing equations.  The evolution of the $\delta \phi_2$ entropy mode is
\begin{align}
\label{delta phi_2 EoM in SRA}
\delta \phi_2'  + \tilde{M}_{22} \delta \phi_2 \approx - \left(\tilde{M}_{23} - Z_{32}\right) \delta \phi_3 + \sum\limits_{m=4}^d \tilde{M}_{2m} \delta \phi_m,
\end{align}
where $Z_{2n}=0$ for $n \ge 4$.  Similarly, for the $\delta \phi_n$ mode, where $n \ge 3$, the sourcing equation is
\begin{widetext}
\begin{align}
\label{delta phi_n EoM slow-roll}
\delta \phi_n' + \tilde{M}_{nn} \delta \phi_n \approx - & \left(\tilde{M}_{n,n-1} + Z_{n,n-1}\right)  \delta \phi_{n-1} - \left(\tilde{M}_{n,n+1}-Z_{n+1,n}\right) \delta \phi_{n+1} + \sum\limits_{m=2,|n-m| \ge 2}^{d} \tilde{M}_{nm} \delta \phi_m.
\end{align}
\end{widetext}
This equation follows from the fact that $\mathbf{Z}$ is skew-symmetric with $Z_{mn}=0$ when $|m-n| \ge 2$.

\subsection{Mode Evolution in the Absence of Sourcing}
\label{no sourcing}

In the remainder of Section \ref{Perbs}, we analyze how the geometry and kinematics of inflation dictates the superhorizon evolution of modes. 

First, consider what happens in the absence of sourcing.  For $\delta \phi_n$ to be unsourced, Eqs. (\ref{delta phi_1 exact EoM}), (\ref{delta phi_2 EoM in SRA}), and (\ref{delta phi_n EoM slow-roll}) show that $\mathbf{e}_n$ must not be turning and the effective mass matrix coefficients $\tilde{M}_{nm}$ must vanish for all $m \neq n$.  When these conditions are met, the $\delta \phi_n$ mode is unsourced and its evolution is governed solely by $\tilde{M}_{nn}$.  In the past, the mode amplitude decay has sometimes been modeled as being approximately proportional to $e^{-\tilde{M}^*_{nn} N_*}$, where $\tilde{M}_{nn}^*$ is the value of the effective mass at horizon exit and $N_*$ is the number of e-folds since the mode exited the horizon.  But as we discussed in \cite{PetersonAndTegmark-2010} for the case of two-field inflation, this assumption often leads to large inaccuracies of up to a few orders of magnitude in estimating the mode amplitudes and the spectra (see \cite{PetersonAndTegmark-2010} and references therein.)  Hence, to accurately model the behavior of the unsourced mode, the expression
\begin{align}
\label{delta phi_n intrinsic EoM}
\delta \phi_n (N) =  \delta \phi_n (N_1) \, e^{- \int_{N_1}^{N} \tilde{M}_{nn}(N_2) dN_2}
\end{align}
should be used. That is, the integral of the effective mass for the $n$th mode, $\tilde{M}_{nn}$, most accurately gives that mode's relative change in the amplitude.

Now the effective mass $\tilde{M}_{nn}$ depends on two quantities: the covariant Hessian of the inflationary potential, $M_{nn} \equiv \nabla_n \nabla_n \ln V$ and $\frac{1}{3} R_{nn}$, where $\mathbf{R}$ depends on the curvature tensor of the field manifold contracted with two field velocity vectors, as shown in Eq. (\ref{R defn}).  Since both are geometric quantities, this means that we can predict the behavior of $\delta \phi_n$ by determining the geometry of the inflationary potential and field metric.   Take first the covariant Hessian of the inflationary potential, $M_{nn}$.  If the potential $\ln V$ is covariantly concave up along the $\mathbf{e}_n$ direction---$\nabla_n \nabla_n \ln V >  0$---this causes $\delta \phi_n$ to decay.  Conversely, if the potential is concave down along the $\mathbf{e}_n$ direction,  $\delta \phi_n$ will grow.  A well-known example of this behavior is the adiabatic mode in single-field inflation, which is often likened to a ball rolling down a hill that speeds up or slows down depending on the concavity of its path.  A second example is the $\delta \phi_2$ mode in two-field inflation; if the two-dimensional field trajectory lies in a valley (concave up), then $\delta \phi_2$ decays, but if it lies along a hill (concave down) in $\ln V$, then $\delta \phi_2$ grows in amplitude.  

The second geometrical quantity involved, the curvature term $R_{nn}$, involves the contraction of the Riemann tensor of the field metric with two instances each of $\mathbf{e}_1$ and $\mathbf{e}_n$.   Geometrically, it represents $2\epsilon$ times the $\mathbf{e}_n$ component of the failure of $\mathbf{e}_1$ to be parallel-transported around a closed loop defined by the directions $\mathbf{e}_1$ and $\mathbf{e}_n$.  Therefore, its contribution to the mode sourcing can in principle be determined from the geometry of the field metric and the potential.  If this deviation from parallel transport of $\mathbf{e}_1$ results in a positive component along the $\mathbf{e}_n$ direction, then this causes $\delta \phi_n$ to grow; conversely, a negative value causes $\delta \phi_n$ to decay.  For example, in two-field inflation, since $R_{22}$ is proportional to negative $\epsilon$ times the Ricci scalar of the field manifold, if the field manifold is locally elliptical, this will cause $\delta \phi_2$ to decay, while a locally hyperbolic surface will cause $\delta \phi_2$ to grow.  Note that if both $M_{nn}$ and $R_{nn}$ have the same sign, they will partially negate each other.   Therefore, we can view the effective mass as some sort of measure of the ``net curvature'' or geometry of the inflationary Lagrangian along the single direction specified by $\mathbf{e}_n$.

\subsection{Mode Conservation Laws}
\label{mode cons laws}

The corollary of Eq. (\ref{delta phi_n intrinsic EoM}) is that when $\delta \phi_n$ is unsourced, the quantity
\begin{align}
\label{adiabatic conservation law}
\delta \phi_n e^{\int \tilde{M}_{nn} dN}
\end{align}
is conserved in the superhorizon limit. 
For example, in single-field inflation, the $\delta \phi_1$ mode is automatically unsourced, and hence the quantity
\begin{align}
\label{delta phi_1 intrinsic EoM}
\delta \phi_1 e^{\int_* \tilde{M}_{11} dN} \propto \frac{\delta \phi_1}{v}
\end{align}
is conserved.
In inflation with two effective fields, the entropy mode $\delta \phi_2$ is unsourced, so the quantity
\begin{align}
\delta \phi_2 e^{\int_* \tilde{M}_{22} dN}
\end{align}
is conserved, allowing one to find a semianalytic expression for $\delta \phi_1$ without needing to solve a set of coupled equations \cite{PetersonAndTegmark-2010}.  Therefore, it follows that \textit{Eq. (\ref{adiabatic conservation law}) is the multifield generalization of the well-known single-field conservation law for the adiabatic mode in Eq. (\ref{delta phi_1 intrinsic EoM}).}  Eq. (\ref{adiabatic conservation law}) endows each of the $d$ modes with a conservation equation that holds whenever $\delta \phi_n$ is not sourced---which happens when $\mathbf{e}_n$ is not turning and the covariant Hessian of $\ln V$ has no off-diagonal components along the direction $\mathbf{e}_n$.   So there are up to $d$ potential conserved quantities related to the modes. 

Understanding a mode's effective mass is not just important in the absence of sourcing, but also when sourcing is important.  For example, consider two-field inflation where the entropy mode $\delta \phi_2$ sources the adiabatic mode. If the effective mass of the entropy mode $\delta \phi_2$ is significantly large and positive, then the entropy mode will decay, thereby reducing the ability of the entropy mode to source the adiabatic mode.  Taking the limit where the effective entropy mass is very large and positive, we can neglect the entropy mode, and the evolution of the adiabatic mode is effectively single-field.  In the opposite limit, if the effective entropy mass is large and negative, the entropy mode will grow rapidly, resulting in much stronger sourcing.  And this later scenario is also more likely to result in large non-Gaussianity (see \cite{PetersonAndTegmark-2010b}).  Extrapolating to general multifield inflation, we expect that the size and magnitude of the effective mass $\tilde{M}_{nn}$ plays a significant role in determining how much $\delta \phi_n$ influences the evolution of $\delta \phi_{n-1}$.  In particular, when the effective mass for the $n$th mode is very large and positive, the scenario is likely to behave like a multifield scenario with $n-1$ effective fields.  In the particular case where all but one of the fields have a large and positive effective mass, the scenario can effectively be treated as single-field with a reduced potential \cite{YamaguchiAndYokoyama-2005}.

\subsection{Sourcing in the Special Case of Quadratic Potentials with Canonical Kinetic Terms}
\label{quad potls}

Before we consider sourcing in the general case, we first consider the mode interactions in the special case of quadratic potentials with canonical kinetic terms.  It turns out that the mode interactions simplify greatly in this scenario and that the results can be used as a reference for comparing other inflationary Lagrangians.

For quadratic potentials with canonical kinetic terms, the effective mass matrix simplifies to 
\begin{align}
\label{quad M}
\tilde{M}_{ij} = M_{ij} =  \partial_i \partial_j \ln V.
\end{align}
and the potential satisfies
\begin{align}
\label{quadratic potl defn}
\partial_i \partial_j \partial_k V = 0
\end{align}
for all $i, j,$ and $k$.  
Therefore, repeatedly taking the derivative of Eq. (\ref{quad M})  and using Eq. (\ref{quadratic potl defn}) gives
\begin{align}
\label{DM/dN}
\frac{D \mathbf{M}}{dN} & = 2\epsilon \left(\mathbf{M} + \boldsymbol{\phi}'  \boldsymbol{\phi}'^\dag\right) - \frac{D}{dN} \left(\boldsymbol{\phi}' \boldsymbol{\phi}'^\dag\right) \nonumber \\
\frac{D^2 \mathbf{M}}{dN^2} & = \left(4\epsilon^2 + 2 \boldsymbol{\phi}' \cdot \boldsymbol{\eta} \right) \left(\mathbf{M} + \boldsymbol{\phi}' \boldsymbol{\phi}'^\dag\right) - \frac{D^2}{dN^2} \left(\boldsymbol{\phi}' \boldsymbol{\phi}'^\dag\right) \nonumber \\
& \, ... \\
\frac{D^n \mathbf{M}}{dN^n} & = V \left(\frac{D^n V^{-1}}{dN^n}\right) \left(\mathbf{M} + \boldsymbol{\phi}'  \boldsymbol{\phi}'^\dag\right) - \frac{D^n}{dN^n} \left(\boldsymbol{\phi}' \boldsymbol{\phi}'^\dag\right). \nonumber
\end{align}

We use the above results to show that the turn rate matrix for quadratic potentials with canonical kinetic terms can be expressed solely in terms of coefficients of the mass matrix in the slow-roll limit.   First, for all inflationary scenarios, the slow-roll equation for the covariant acceleration $\boldsymbol{\eta}$ implies that
\begin{align}
\label{Z_21 quad}
Z_{21}  = - M_{21},
\end{align} 
which follows from projecting Eq. (\ref{eta slow-roll v1}) onto the basis vector $\mathbf{e}_2$.  Differentiating Eq. (\ref{eta slow-roll v1}) and projecting it onto $\mathbf{e}_n$, where $n \ge 3$, gives
\begin{align}
\label{jerk quad}
\xi_n = - M_{n2} \eta_2 - \left(\frac{DM}{dN}\right)_{n1} v,
\end{align}
where we have used the facts that $M_{n1} = 0$ and $\eta_n = 0$ for all $n \ge 3$.  But for quadratic potentials with trivial field metrics, $\left(\frac{DM}{dN}\right)_{n1}$ also equals 0 in virtue of Eq. (\ref{DM/dN}) and the facts that $M_{n1} = 0$ and that $\chi^{(m)}_n = 0$ for $m < n$.  Using this result in Eq. (\ref{jerk quad}), it therefore follows that
\begin{align}
\label{Z_32 quad}
Z_{32}  = - M_{32} 
\end{align}
and 
\begin{align}
\label{M_n2 quad}
M_{n2}  = 0   \, \, \, \, \, \, \, \textrm{for} \, \, n \ge 4.
\end{align}
Similarly, differentiating Eq. (\ref{eta slow-roll v1}) a second time, projecting it onto $\mathbf{e}_n$ where $n \ge 4$, and using Eq. (\ref{M_n2 quad}) gives
\begin{align}
\label{jerk quad}
\chi^{(4)}_n = - M_{n3} \xi_3 - 2\left(\frac{DM}{DN}\right)_{n2} \eta_2 - \left(\frac{D^2M}{DN^2}\right)_{n1} v.
\end{align}
But by Eq. (\ref{DM/dN}), $\left(\frac{DM}{DN}\right)_{n2}$  and $\left(\frac{D^2M}{DN^2}\right)_{n1}$ vanish for $n \ge 4$, yielding
\begin{align}
\label{Z_43 quad}
Z_{43}  & = - M_{43}  \\
M_{n3}  & = 0  \, \, \, \, \, \, \, \textrm{for} \, \, n \ge 5 \nonumber
\end{align}
Repeating this series of steps, we find that for this special class of scenarios
\begin{align}
\label{quadratic special relation 1}
Z_{n+1,n}  = - M_{n+1,n}
\end{align} 
and 
\begin{align}
\label{quadratic special relation 2}
M_{mn} & = 0  \, \, \, \, \, \, \, \textrm{for} \, \, |m - n| \ge 2 \\
\left(\frac{D^p M}{DN^p}\right)_{mn} & =  0   \, \, \, \, \, \, \, \textrm{for} \, \, |m - n| \ge p + 1, \nonumber
\end{align}
where $p \ge 1$.
 
Eq. (\ref{quadratic special relation 1}) shows that the turn rate matrix can be expressed entirely in terms of coefficients of the mass matrix.  The rate at which the $\mathbf{e}_n$ basis vector turns into the direction of the $\mathbf{e}_{n+1}$ is given simply by $-M_{n+1,n}$.  Substituting this result into Eq. (\ref{delta phi_n EoM slow-roll}), the mode sourcing equation for all $d$ modes reduces to
\begin{align}
\label{quadratic sourcing eqtn}	
\delta \phi_n' + M_{nn} \, \delta \phi_n \approx  2 Z_{n+1,n} \, \delta \phi_{n+1}.
\end{align}
Therefore, whenever the Lagrangian consists of a quadratic potential and canonical kinetic terms, the $\delta \phi_n$ mode is sourced only by the $\delta \phi_{n+1}$ mode; the other $d-2$ do not influence $\delta \phi_n$, so only a single interaction needs to be considered for each mode.   Moreover, the $\delta \phi_n$ mode is sourced only when the $\mathbf{e}_n$ basis vector rotates into the $\mathbf{e}_{n+1}$ direction.  This provides a very simple way to understand this special class of Lagrangians in terms of the geometry/kinematics of inflation.  It also explains why these scenarios are the simplest to solve: every mode obeys an equation of motion that is identical in form to the adiabatic mode.   Mathematically, the solution for the $n$th mode becomes 
\begin{align}
\delta \phi_n \approx  \delta \phi_n^* & \, e^{-\int_{N_*}^{N} M_{nn} dN_1} \\ & + \int_{N_*}^{N} 2 Z_{n+1,n} \delta \phi_{n+1} \, e^{- \int_{N_*}^{N_1} M_{nn} dN_2} \, dN_1.  \nonumber
\end{align}
In principle, one could solve the above series of integrals.  Finally, since there is no sourcing when $Z_{n+1,n}=0$, the number of kinematical basis vectors that are changing direction inversely indicates the number of conserved mode quantities. 

The results for quadratic potentials with trivial field metrics are not just interesting in and of themselves, but they provide a critical vantage point from which to understand the mode interactions in general multifield inflation, as we will show in the next section.

\subsection{Sourcing in the General Case}
\label{sourcing}

Finally, we consider entropy mode sourcing in the general case.  We start by discussing the three types of terms that can give rise to sourcing effects.   Then we discuss how general multifield inflation differs from the canonical quadratic case and how various order covariant derivatives of the potential affect the mode interactions.

According to Eq. (\ref{delta phi_n EoM slow-roll}), sourcing effects can arise from:
\begin{enumerate}
\item Off-diagonal terms in the mass matrix 
\item Any nontrivial geometry of the field manifold  
\item The kinematical basis vectors changing direction 
\end{enumerate}
We will consider each sourcing effect in turn.

The first type of sourcing effect arises from off-diagonal terms in the mass matrix, which is defined as the covariant Hessian of $\ln V$.  These off-diagonal terms are measures of the coupling between fields in the potential and of whether this coupling results in a higher or lower potential energy state.  But these terms can also be viewed as geometric effects because $M_{nm}$ represents how much the $n$th component of the covariant derivative of $\ln V$ varies along the $\mathbf{e}_{m}$  direction.  Therefore the shape of the potential provides insight into this type of sourcing effect.   If the coupling term $M_{nm}$ is positive, then $\delta \phi_{m}$ will cause $\delta \phi_n$ to decay; otherwise, if it is negative, it will increase the amplitude of $\delta \phi_n$.  Interestingly, since the mass matrix is symmetric, a nonzero $M_{n,n+1}$ leads to parallel sourcing effects; for example,  a negative value for $M_{n,n+1}$ will cause both the $\delta \phi_n$ and $\delta \phi_{n+1}$ modes to grow.\footnote{Again, when working in the classical picture, our statements are with respect to positive field fluctuations; it is straightforward to extrapolate to other cases.}   

The second type of sourcing effect arises from the curvature matrix $\mathbf{R}$.  As explained earlier, the form of the kinetic terms in the inflationary Lagrangian can be represented through a field metric, and this metric can be viewed as inducing a field manifold.   If the field manifold has nontrivial geometry, then the Riemann curvature tensor will be nonzero, and this will be manifested in the form of a nonzero symmetric curvature matrix $R^a_d \equiv 2\epsilon R^a_{bcd} e_1^b e_1^c$.  Specifically, if the $\mathbf{e}_m$ component of the failure of $\mathbf{e}_1$ to be parallel transported around the closed loop defined by $\mathbf{e}_1$ and $\mathbf{e}_n$ is nonzero, then the curvature matrix will cause $\delta \phi_m$ to source $\delta \phi_n$.  Since the curvature matrix can be factored into $\epsilon$ times a term involving the Riemann tensor, this term technically combines geometric and kinematical effects; so when all else is equal, the impact of noncanonical terms on the mode sourcing tends to be greatest at the end of inflation and whenever else the field speed is large.  Now like the mass matrix, since the curvature matrix is symmetric, a positive value for a given curvature matrix coefficient will cause both the $\delta \phi_m$ and $\delta \phi_{n}$ modes to grow. Note that in comparison to the sourcing effects due to the mass matrix coefficients, the curvature matrix appears in the equation of motion with the opposite sign.  Thus, we may view the mass matrix and curvature matrix as representing sourcing effects due to the geometry of the Lagrangian, with the mass matrix primarily corresponding to the potential and the curvature matrix to the field metric.

\begin{table}[t]
\centering
\renewcommand{\arraystretch}{1.5}
\begin{tabular}{|x{0.96in}|x{2.29in}|} \hline
\rule{0cm}{0.55cm} Sourcing Terms & What the Terms Represent \tabularnewline[1.4ex] \hline
\multicolumn{2}{|c|}{$\, $}\tabularnewline [-3.7ex] \hline 
\rule{0cm}{0.55cm} $M_{nm}$ & Covariant Hessian of Potential (geometry of potential)  \tabularnewline [1.4ex] \hline
\rule{0cm}{0.55cm} $R_{nm}$ & Riemann Tensor of Field Manifold, $\epsilon$ (geometry of field manifold, kinematics) \tabularnewline [1.4ex] \hline
\rule{0cm}{0.55cm} $Z_{n+1,n}$, $Z_{n,n-1}$ & Turn Rate of $\mathbf{e}_n$ (kinematics)  \tabularnewline [1.4ex]  \hline
\multicolumn{2}{c}{$\, $}
\end{tabular}
\caption{The three types of sourcing effects in the mode sourcing equation for $\delta \phi_n$ and what each set of terms effectively represents.  More detailed explanation of the terms and their impact on mode sourcing is given in the text below.}
\end{table}

The third and last kind of sourcing effect is purely a kinematical effect---a direct consequence of the kinematical basis vectors changing direction.  Importantly, the coefficients of turn rate matrix allow $\delta \phi_n$ to be sourced by only two modes: $\delta \phi_{n-1}$ and $\delta \phi_{n+1}$.  Consider first the term  $Z_{n+1,n} \delta \phi_{n+1}$ in Eq. (\ref{delta phi_n EoM slow-roll}).  The kinematical term $Z_{n+1,n}$ represents how quickly the $\mathbf{e}_n$ basis vector is turning into the direction of the $\mathbf{e}_{n+1}$ basis vector.  Since $Z_{n+1,n}$ is always non-negative, this turning will always cause $\delta \phi_n$ to grow.   And the faster $\mathbf{e}_n$ is turning into the $\mathbf{e}_{n+1}$ direction, the more $\delta \phi_{n+1}$ sources $\delta \phi_n$.  This sourcing effect can be interpreted physically as follows:  the direct rotation of the kinematical basis vectors causes what was once a $\delta \phi_{n+1}$ mode to be partially converted into a $\delta \phi_n$ mode.   The other kinematical sourcing term, $-Z_{n,n-1} \delta \phi_{n-1}$, can be understood similarly.  However, this term causes $\delta \phi_n$ to shrink in magnitude, which can be explained by the fact that $\delta \phi_n$ is being partially converted into $\delta \phi_{n-1}$ by the rotation of basis vectors.   The anti-symmetry of the turn rate matrix neatly encapsulates these antithetical kinematic effects.  

Though we have dubbed the third type of sourcing a kinematical effect, the question naturally arises as to whether the kinematics can be directly related back to the geometry of the Lagrangian.  In the case of a quadratic potential with canonical kinetic terms, we saw that this is the case and the turn rates can be expressed neatly in terms of the mass matrix coefficients.  But in general this is not true, and the turn rates involve more complex combinations of the various $n$th-order covariant derivatives of $\ln V$.   Hence it is often easiest to view the effects from the turn rate matrix as kinematical effects, rather than a complicated combination of geometric effects.   What is different in the general case of multifield inflation is that $\boldsymbol{\nabla}^n V \neq \mathbf{0}$ for $n \ge 3$ and $\mathbf{R} \neq \mathbf{0}$, producing additional terms in the mode sourcing equations.  This can be seen by starting with the slow-roll expansion for the $n$th kinematical vector,
\begin{align}
\label{chi binomial thm result}
\boldsymbol{\chi}^{(n+2)} =- \sum\limits_{m=0}^n \binom{n}{m} \left[\left(\frac{D}{dN}\right)^{m} \boldsymbol{M}\right] \boldsymbol{\chi}^{(n-m+1)},
\end{align}
which follows from differentiating Eq. (\ref{eta slow-roll v1}) for $\boldsymbol{\eta}$ a total of $n$ times.  For instance, the jerk is
\begin{align}
\boldsymbol{\xi} & = - \mathbf{M} \boldsymbol{\eta} - \frac{D\boldsymbol{M}}{dN} \boldsymbol{\phi}' 
\end{align}
Since $\xi$ has only three nonzero components in the kinematical basis, projecting the jerk onto the basis vectors gives
\begin{align}
\xi_3 & = - M_{32}  - \left(\frac{DM}{dN}\right)_{31} v \\
\left(\frac{DM}{dN}\right)_{n1} & = - M_{n2} \, Z_{21}  \, \, \, \, \, \, \, \textrm{for} \, \, n \ge 4. \nonumber
 \end{align}
 Notice the presence of the extra term $ \left(\frac{DM}{dN}\right)_{n1}$ where $n \ge 3$ in each of the two equations above.  It no longer vanishes because $\boldsymbol{\nabla}^3 V \neq 0$ and instead it equals 
  \begin{align}
 \left(\frac{DM}{dN}\right)_{n1} v =  \mathbf{e}_n \cdot \frac{\boldsymbol{\phi'}^\dag \boldsymbol{\phi'}^\dag \boldsymbol{\nabla}  \boldsymbol{\nabla}  \boldsymbol{\nabla} V}{V},
 \end{align}
causing the turn rate $Z_{32}$ to no longer equal $-M_{32}$: 
\begin{align}
Z_{32} & = - M_{32} + \frac{1}{M_{21}} \mathbf{e}_3 \cdot \left(\frac{\boldsymbol{\phi'}^\dag \boldsymbol{\phi'}^\dag \boldsymbol{\nabla}  \boldsymbol{\nabla}  \boldsymbol{\nabla} V}{V}\right) \\
\end{align}
for $M_{21} \neq 0$.  So both $\left(\frac{DM}{dN}\right)_{n1}$ and $M_{n2}$ no longer vanish for large enough $n$ but are interrelated: 
\begin{align}
\left(\frac{DM}{dN}\right)_{n1} & = - M_{n2} \, Z_{21}  \, \, \, \, \, \, \, \textrm{for} \, \, n \ge 4.
 \end{align}
Thus in comparison to quadratic potentials with canonical kinetic terms, $Z_{32}$ picks up extra terms that depend on the third covariant derivative of $V$.  Similarly, one can show that the next highest turn rate in the series is
\begin{widetext}
\begin{align}
Z_{43} & = - M_{43} + \frac{1}{\xi_3} \left[M_{42} (-\xi_2 + 2\epsilon \eta_2) - \mathbf{e}_4 \cdot \left(\frac{\boldsymbol{\phi'}^\dag \boldsymbol{\phi'}^\dag \boldsymbol{\phi'}^\dag \boldsymbol{\nabla}  \boldsymbol{\nabla}  \boldsymbol{\nabla}  \boldsymbol{\nabla} V}{V}\right)\right] 
\end{align}
\end{widetext}
 for $\xi_3 \neq 0$.  The result here differs from the simple quadratic case by the presence of terms that depend on the third and fourth covariant derivatives of $V$, including the term $M_{42}$.  In general, one can show that 
\begin{align}
\label{Z relate to M}
Z_{n+1,n} \approx - M_{n+1,n} + \textrm{higher-order corrections}
\end{align}
where the ``corrections'' vanish for $n=1$ but otherwise depend on the higher-order covariant derivatives $\boldsymbol{\nabla}^{p} V$ up to order $p=n+1$.  Interestingly, plugging Eq. (\ref{Z relate to M}) into the entropy mode equation (\ref{delta phi_n EoM slow-roll}) tells us that the sourcing of $\delta \phi_n$ by $\delta \phi_{n-1}$ is controlled by these corrections arising from higher-order covariant derivatives:
\begin{widetext}
\begin{align}
\delta \phi_n' + \tilde{M}_{nn} \delta \phi_n \approx - & \left(\mathrm{corrections}\right)  \delta \phi_{n-1} + \left(2Z_{n+1,n} + \mathrm{corrections} \right) \delta \phi_{n+1} - \sum\limits_{m=2,|n-m| \ge 2}^{d} M_{nm} \delta \phi_m + \frac{1}{3} \sum\limits_{m=2,m \neq n}^{d} R_{nm} \delta \phi_m.
\end{align}
\end{widetext}
Similarly, we can view the sourcing of $\delta \phi_n$ by $\delta \phi_{n+1}$ to be controlled by a term that is twice the turn rate $Z_{n+1,n}$, plus corrections from higher-order covariant derivatives of $V$.   

This results in a very interesting and useful way to view the interactions among modes: the interactions can essentially be divided into sourcing effects shared in common with canonical quadratic models ($Z_{n+1,n}$ terms) and sourcing effects arising from deviations from this fundamental Lagrangian (the higher-order derivatives of $V$ and the corrections from the field metric, $R_{nm}$).   Taylor expansion of the inflationary potential with an understanding of the relative size of the various order terms in the expansion can therefore indicate how much each term $Z_{n+1,n}$ differs from $-M_{n+1,n}$ and hence the degree to which the scenario differs from the canonical quadratic case, as we illustrated above.   We advocate this novel approach as a powerful prescription for exploring how differences in inflationary Lagrangians translate into differences in mode dynamics.

\section{Spectral Observables}
\label{CosmicObs}

Sections \ref{Background Fields} and \ref{Perbs} explored how the kinematics and geometry of the inflationary potential and the field manifold determine the evolution of modes.  In this section, we connect these results to the cosmic observables.   Since most of these connections follow straightforwardly from our discussion of mode sourcing in Sections \ref{no sourcing} - \ref{sourcing}, here we focus on how the inflationary geometry and kinematics determine the effective number of inflationary fields in Section \ref{dimension} and how this is reflected in the cosmic power spectra, bispectrum, and trispectrum in Sections \ref{Pwr Spectra}-\ref{nonGauss}.    In the process of doing so, we introduce a new cosmic multifield observable that can potentially distinguish two-field models from models with three or more fields (Section \ref{isocurv}), and we present a new multifield consistency relation (Section \ref{nonGauss}).

\subsection{Effective Number of Fields}
\label{dimension}

We define the effective number of fields or \textit{dimension} of inflation to be the minimum number of fields necessary to adequately describe both the background and perturbed solutions for inflation across the distance scales of interest.   

To represent the background solution, the minimum number of fields is the same as the number of fields needed to reproduce all the kinematical vectors, as defined in Eq. (\ref{kine vecs}).  This corresponds to the number of basis vectors needed to span the space defined by the kinematical vectors.  Because of the way we constructed the kinematical basis vectors in Eq. (\ref{basis vecs}), the dimension also equals one plus the number of kinematical basis vectors that are changing direction.   If no kinematical basis vectors are changing direction, then inflation is single-field.  If only the $\mathbf{e}_1$ basis vector is changing direction, then the inflationary scenario has two effective field degrees of freedom; this is the single turn rate that characterizes the dynamics of two-field inflation.  Similarly, the number of unique nonzero components of the turn rate matrix determines the dimension of multifield inflation.   In geometric terms, inflationary scenarios with canonical kinetic terms will produce trajectories lying along a line if single-field, a plane if two-field, and so on.   This is illustrated in Figure \ref{TubesFig}.   Extrapolating to noncanonical kinetic terms, the modification is that the geometry of the background trajectory will be determined with respect to parallel transport of the kinematical vectors.  In either case, the number of basis vectors that change direction along the trajectory provides an intuitive way to understand the effective dimension of the background solution.

The number of fields representing the perturbed solution is more difficult to determine.  We define the minimum number of fields to describe the perturbed solution as the number needed to reproduce solutions for $\delta \phi_1$ (the adiabatic mode) and $\delta \phi_2$ (the first entropy mode).  In the case where the $\mathbf{e}_1$ basis vector never turns, the adiabatic mode is never sourced and the quantity $\frac{\delta \phi_1}{v}$ is conserved in the superhorizon limit, just like in single-field inflation.  However, the dimension of the perturbed fields can still be more than one if there are two or more fields during inflation and hence a power spectrum of entropy modes. 

In the case where the field trajectory does change direction during inflation, there are two reasons why the effective dimension of the background and perturbed fields do not necessarily coincide.  The first reason is that the curvature matrix $\mathbf{R}$ can couple together the various entropy modes, independently of the turning behavior of the kinematical basis vectors.   Second, in general, it is not true that $Z_{n+1,n} \approx - M_{n+1,n}$, as we showed earlier.  So even if the kinematical basis vector $\mathbf{e}_{n}$ is not turning, a nonzero $M_{n+1,n}$ could still allow the $\delta \phi_{n+1}$ mode to source the $\delta \phi_n$ mode.  Similarly, it is possible for higher-order covariant derivatives of the potential to produce a nonzero turn rate $Z_{n+1,n}$ even if $M_{n+1,n}=0$.  (Of course, for many models, when $Z_{n+1,n} = 0$, it will also be true that $M_{n+1,n} = 0$.)   Therefore, for models with at least two fields, the effective number of field perturbations we need to consider in order to find expressions for $\delta \phi_1$ and $\delta \phi_2$ equals two plus the number of consecutive sourced perturbations when starting at $\delta \phi_3$ and counting upwards in the series of modes.  This follows directly from the series of slow-roll sourcing equations in Eq. (\ref{delta phi_n EoM slow-roll}).  Therefore, the exact same geometric and kinematical quantities that determine the number and strength of sourcing relationships can be used to determine the effective dimension of the perturbed fields.   In particular, scenarios with at least one large positive effective mass $\tilde{M}_{n+1,n+1}$ and/or a negligible turn rate $Z_{n+1,n}$ over all scales of interest are prime candidates for dimensional reduction; such features usually indicate that $\delta \phi_{n+1}$ has a negligible impact on $\delta \phi_n$ and that the series of mode sourcing equations can be truncated after $\delta \phi_n$.   

Based on the above analysis, we take the effective dimension of the perturbed field system, which can be larger than the dimension of the unperturbed system, as the overall effective dimension of a multifield scenario.  Yet although we can assign an overall dimension to each scenario, it is also useful to consider that an inflationary scenario may be broken into multiple phases, with each one defined by a different effective number of fields being active.  
For example, in canonical quadratic models with very different masses for the fields in the potential, there are periods dominated by the dynamics of a single-field, punctuated by periods in which two fields dominate the dynamics.  By understanding that a model with multiple fields can be approximated by a series of scenarios with a much smaller effective dimension---such as a series of single-field and two-field scenarios---we can gain much greater insight into the key features of such models, and they become more computationally tractable.

\subsection{Tensor Power Spectra}
\label{Pwr Spectra}

With these insights, we explore the main spectral observables to see how they reflect the effective dimension of multifield scenarios. 

We start with the power spectra.  The tensor power spectrum is unchanged by the presence of multiple fields and has the form \cite{Starobinski-1979}
\begin{align}
P_T = 8 \left(\frac{H_*}{2\pi}\right)^2,
\end{align}
under the common convention for normalization of the spectrum.  The tensor spectral index represents the scale-dependence of the tensor spectrum and is defined as
\begin{align}
n_T & \equiv \frac{d \ln \mathcal{P}_T}{d \ln k}.
\end{align}
Since $d \ln k = d N$ to first-order in slow-roll, 
\begin{align}
n_T & = - 2 \epsilon_*,
\end{align} 
and $n_T$ depends only on the speed of the field vector and not on any other kinematic or geometric properties of inflation.

\subsection{Transfer Matrix Formalism}
\label{transfer}

The scalar power spectra are typically given in terms of the spectra of curvature and isocurvature perturbations.   The curvature perturbation $\mathcal{R}$ during inflation is related to the adiabatic density mode by \cite{SasakiAndTanaka-1998}
\begin{align}
\mathcal{R} = \frac{\delta \phi_1}{v}.
\end{align}
The isocurvature modes, $\mathcal{S}$, are typically defined in the following gauge-invariant and dimensionless manner \cite{GordonEtAl-2000, WandsEtAl-2002}: 
\begin{align}
\mathcal{S} \equiv \frac{\delta p}{p'} - \frac{\delta \rho}{\rho'}.
\end{align}
Calculating the above quantity reveals that $\mathcal{S}$ depends only on the entropy mode $\delta \phi_2$, up to normalization factors.  Here we choose the normalization factor so that the isocurvature and curvature spectra have equal power at horizon crossing:
\begin{align}
\mathcal{S} \equiv \frac{\delta \phi_2}{v}.
\end{align}  

The relationship between the curvature and isocurvature modes can be described in terms of the transfer matrix formalism \cite{AmendolaEtAl-2001,WandsEtAl-2002}.  In two-field inflation, the transfer matrix formalism represents the evolution of curvature and isocurvature modes as
\begin{align}
\label{transfer matrix}
\left(\begin{array}{c} \mathcal{R} \\ \mathbf{\mathcal{S}} \end{array} \right) = & 
\left(\begin{array}{cc} 1 & T_{\mathcal{RS}} \\ 0 & T_{\mathcal{SS}} \end{array} \right)
\left(\begin{array}{c} \mathcal{R}_* \\ \mathcal{S}_* \end{array} \right),
\end{align}
which follows from the fact that in two-field inflation, the adiabatic mode is sourced by the entropy mode but not vice versa.  The transfer function $T_{\mathcal{RS}}$ represents the sourcing of the curvature modes by the isocurvature modes, while the transfer function $T_{\mathcal{SS}}$ represents the intrinsic evolution of the isocurvature modes.  In general multifield inflation, a collection of entropy modes replaces the single entropy mode represented by $\mathcal{S}$, so the transfer matrix formalism can be generalized as
\begin{align}
\label{transfer matrix multi}
\left(\begin{array}{c} \mathcal{R} \\ \boldsymbol{\frac{\delta \phi_{\perp}}{v}} \end{array} \right) = & 
\left(\begin{array}{cc} 1 & \mathbf{T_{\mathcal{R}\perp}} \\ 0 & \mathbf{T_{\perp \perp}} \end{array} \right)
\left(\begin{array}{c} \mathcal{R}_* \\ \boldsymbol{\frac{\delta \phi_{\perp}^*}{v_*}} \end{array} \right),
\end{align}
where $\boldsymbol{\frac{\delta \phi_{\perp}}{v}}$ is a $d-1$ dimensional vector and the analogous transfer functions are the vector $\mathbf{T_{\mathcal{R \perp}}}$ and the matrix $\mathbf{T_{\perp \perp}}$.  The expression for $\mathbf{T}_{\perp \perp}$ represents the evolution of the entropy mode vector since horizon exit.  But despite the presence of additional entropy modes, it is still true that only the $\delta \phi_2$ entropy mode sources $\delta \phi_1$; this follows from Eq. (\ref{delta phi_1 exact EoM}), which can be rewritten as
\begin{align}
\frac{D\mathcal{R}}{dN} = 2 Z_{21} \mathcal{S}.
\end{align}
So to find the curvature and isocurvature modes at the end of inflation, we need to know how the $d-2$ entropy modes source the $\delta \phi_2$ mode and in turn how the $\delta \phi_2$ mode sources the $\delta \phi_1$ mode.  This can be represented in terms the transfer functions
\begin{align}
\label{multifield TRS v1}
\mathbf{T}_{\mathcal{R}\perp}(N) & \equiv \int_{N_*}^{N} 2 Z_{21}(N_1) \, \mathbf{T}_{\mathcal{S} \perp}(N_1) \, dN_1, \nonumber \\
\mathbf{T}_{\mathcal{S}\perp}(N)  & \equiv  \mathbf{e}_2(N) \cdot \mathbf{T}_{\perp \perp}(N),
\end{align} 
where the time-dependence is indicated explicitly.  

To find an expression for $\mathbf{T}_{\mathcal{\perp \perp}}(N)$, we return to the expression for the evolution of entropy modes in Eq.  (\ref{perp perb EoM}).  From this equation, it follows that
\begin{align}
\label{Tperpperpmatrix}
\mathbf{T}_{\mathcal{\perp \perp}}(N)  & \equiv  \frac{1}{v(N)} \, e^{- \int_{N_*}^{N} \left[\mathbf{\tilde{M}}_{\perp \perp}(N_1) + \mathbf{Z}_{\perp \perp}(N_1) \right] \, dN_1}
\end{align} 
to lowest order in the slow-roll limit.   If no approximate analytic solution for $\mathbf{T}_{\perp \perp}$ can be found, the solution can be estimated using the Magnus series expansion.  According to the Magnus series expansion (see \cite{BlanesEtAl-2008} and references therein), if 
\begin{align}
e^{\boldsymbol{\Omega}(N)} \equiv e^{- \int_{N_*}^{N} \mathbf{A}_1 dN_1}, 
\end{align}
where $\mathbf{A}_1 \equiv \mathbf{A}(N_1)$, then the first three terms in the series expansion are
\begin{widetext}
\begin{align}
\boldsymbol{\Omega}_1 = & - \int_{N_*}^N  \mathbf{A}_1 dN_1, \nonumber \\
\boldsymbol{\Omega}_2 = & \frac{1}{2} \int_{N_*}^N \int_{N_*}^{N_1}  [\mathbf{A}_1, \mathbf{A}_2]  \, dN_2 dN_1, \\
\boldsymbol{\Omega}_3 = & -\frac{1}{3!} \int_{N_*}^N \int_{N_*}^{N_1} \int_{N_*}^{N_2} ( [\mathbf{A}_1, [\mathbf{A}_2, \mathbf{A}_3]] + [\mathbf{A}_3, [\mathbf{A}_2,\mathbf{A}_2] \, ) \, dN_3 \, dN_2 \, dN_1, \nonumber
\end{align}
\end{widetext}
where $[\mathbf{A},\mathbf{B}] \equiv \mathbf{AB} - \mathbf{BA}$ is the matrix commutator of matrices $\mathbf{A}$ and $\mathbf{B}$ and here 
\begin{align}
\mathbf{A}(N)  \equiv  \mathbf{\tilde{M}}_{\perp \perp}(N) + \mathbf{Z}_{\perp \perp}(N).
\end{align}
Fortunately, the Magnus expansion for Eq. (\ref{Tperpperpmatrix}) simplifies because $\boldsymbol{\tilde{M}}$ and $\mathbf{Z}$ are symmetric and anti-symmetric, respectively, so their commutator vanishes.  It follows that only the commutators of each matrix with itself at different time points remain and Eq. (\ref{Tperpperpmatrix}) can be decomposed as
\begin{align}
\label{Tperpperpmatrix}
\mathbf{T}_{\mathcal{\perp \perp}}(N)  & \equiv  \frac{1}{v(N)} \, e^{- \int_{N_*}^{N} \mathbf{\tilde{M}}_{\perp \perp}(N_1) \, dN_1} \, e^{- \int_{N_*}^{N} \mathbf{Z}_{\perp \perp}(N_1)  \, dN_1},  
\end{align} 
with the Magnus expansion applied to each matrix exponential separately.  Additional gains in reducing the computational complexity of $\mathbf{T}_{\mathcal{\perp \perp}}(N)$ are possible whenever the space of entropy modes can be dimensionally reduced.   This can be done whenever the series of kinematical mode sourcing relations can be truncated, as discussed in detail in Sections \ref{no sourcing}-\ref{sourcing} and \ref{dimension}.

The dependence of the transfer functions on the geometry and kinematics of inflation follow from our discussions of the mode sourcing relations in Sections \ref{no sourcing}-\ref{sourcing}; however, we provide a few examples here for illustration.   The transfer function $T_{\mathcal{R}\perp}$ depends on the turn rate of the background trajectory times the transfer function $\mathbf{T}_{\mathcal{S}\perp}$, a vector function representing how much the $\delta \phi_2$ mode is sourced by the other $d-2$ entropy modes modulo a factor of $v$.   For example, if the $\mathbf{e}_2$ basis vector is rapidly turning into the $\mathbf{e}_3$ direction while $\mathbf{e}_1$ also turns signficantly, then $\delta \phi_2$ will be strongly sourced by $\delta \phi_3$, causing a boost in the amplitude of both transfer functions.  As a second example, if the field trajectory rolls along a ridge in the potential while negligibly turning, then the $\delta \phi_2$ mode will dramatically grow in amplitude, causing a boost in $\mathbf{T}_{\mathcal{S}\perp}$ but only a small increase in $\mathbf{T}_{\mathcal{R}\perp}$.  As a third example, if a strong negative curvature $R_{32}$ arises from the kinetic terms in the Lagrangian and dominates the dynamics of the $\delta \phi_2$ and $\delta \phi_3$ modes, both modes will decay, thereby reducing $\mathbf{T}_{\mathcal{S}\perp}$ and blunting the sourcing function $\mathbf{T}_{\mathcal{R}\perp}$.    Thus, we emphasize that one can understand how the Lagrangian translates into the spectral observables by studying the mode sourcing in detail.

\subsection{Curvature Spectrum}

Now we find the scalar spectra in terms of the transfer matrix formalism.  The beauty of the transfer matrix formalism is that the multifield spectra follow from the single-field results but with the promotion of the transfer functions from scalars to vectors.

For the curvature spectrum, we make the canonical assumption that following inflation, curvature modes are conserved on superhorizon scales, and so the density and curvature spectra are equivalent up to factors of $O(1)$. Employing the transfer matrix formalism, the curvature power spectrum at the end of inflation \cite{NibbelinkAndVanTent-2001,AmendolaEtAl-2001,WandsEtAl-2002} can be written as
\begin{align}
\label{pwr spec}
\mathcal{P}_{\mathcal{R}} = \left(\frac{H_*}{2\pi}\right)^2 \frac{1}{2\epsilon_*} \left(1 + \left\vert\mathbf{T_{\mathcal{R}\perp}}\right\vert^2\right),
\end{align}
where it is understood that the function $\mathbf{T_{\mathcal{R}\perp}}$ is evaluated at the end of inflation.\footnote{We take the end of inflation to correspond to $\epsilon = 1$, but in principle, another endpoint may be chosen instead.}  Eq. (\ref{pwr spec}) shows that the curvature spectrum at the end of inflation equals the curvature spectrum at horizon exit plus an enhancement due to sourcing of the density mode, $\mathbf{T}_{\mathcal{R}\perp}$.

To determine how the effective number of fields is reflected in the spectra, we define a new unit vector
\begin{align}
\label{e_R unit vec}
\mathbf{e}_{\mathcal{R}}  \equiv \frac{\mathbf{T}_{\mathcal{R}\perp}}{|\mathbf{T}_{\mathcal{R}\perp}|}
\end{align}
and the scalar quantity
\begin{align}
\label{TRperp mag}
T_{\mathcal{R}\perp} \equiv |\mathbf{T}_{\mathcal{R}\perp}|.
\end{align}
The unit vector $\mathbf{e}_{\mathcal{R}}$ necessarily lies in the ($d-1$)-dimensional subspace spanned by the kinematical basis vectors $\mathbf{e}_2^*,\mathbf{e}_3^*,...,\mathbf{e}_d^*$, where again $*$ represents that a quantity is evaluated at horizon exit.  If inflation has two effective fields, then $\mathbf{e}_{\mathcal{R}} = \mathbf{e}_2^*$; however, if inflation has more than two effective fields, then $\mathbf{e}_{\mathcal{R}} \neq \mathbf{e}_2^*$.  Moreover, one plus the number of nonzero components of $\mathbf{e}_{\mathcal{R}}$ in the kinematical basis gives the effective number of fields.  Therefore to probe the number of effective fields during inflation, we need to obtain information on the number of nonzero components of $\mathbf{e}_{\mathcal{R}}$.  But by Eq. (\ref{TRperp mag}), the curvature power spectrum for general multifield inflation can be rewritten as
\begin{align}
\label{two-field curv spec}
\mathcal{P}_{\mathcal{R}} = \left(\frac{H_*}{2\pi}\right)^2 \frac{1}{2\epsilon_*} \left(1 + T_{\mathcal{R}\perp}^2\right),
\end{align}
which eliminates $\mathbf{e}_{\mathcal{R}}$ from the expression and renders  Eq. (\ref{two-field curv spec}) identical in form to the corresponding expression for two-field inflation \cite{PetersonAndTegmark-2010}.   This  means that the curvature spectrum provides no insight into the number of fields during inflation. 

However, combining the curvature and tensor spectra together does reveal whether inflation is single-field or multifield, as is well-known.  For single-field inflation, we necessarily have $T_{\mathcal{R}\perp} = 0$ and therefore the tensor-to-scalar ratio $r_T$, defined by
\begin{align}
r_T \equiv \frac{\mathcal{P}_{\mathcal{R}}}{\mathcal{P}_T},
\end{align}
produces the single-field consistency relation 
\begin{align}
\label{single-field rT consistency reln}
r_T = -8 n_T.
\end{align}
In multifield inflation, however, the ratio satisfies the upper bound \cite{WandsEtAl-2002,PetersonAndTegmark-2010}:
\begin{align}
\label{rT inequality}
r_T = -8n_T \cos^2 \Delta_N \le -8n_T,
\end{align}
where 
\begin{align}
\tan \Delta_N = T_{\mathcal{R}\perp}.
\end{align}
Therefore, if the upper bound in Eq. (\ref{rT inequality}) is not saturated, then inflation is multifield.

As an aside, the multifield curvature power spectrum can also be given in terms of the $\delta N$ formalism.  Under the $\delta N$ formalism, correlators of $\mathcal{R}$ can be written in terms of covariant derivatives of $N$, so the curvature power spectrum can be written as \cite{SasakiAndStewart-1995}
\begin{align}
\label{pwr spec in delta N formalism}
\mathcal{P}_{\mathcal{R}} = \left(\frac{H_*}{2\pi}\right)^2 \left\vert \boldsymbol{\nabla} N \right\vert^2,
\end{align}
where $\boldsymbol{\nabla} N$ is the covariant derivative of the number of e-folds of inflation.
By comparing Eqs. (\ref{pwr spec}) and (\ref{pwr spec in delta N formalism}) and using that $\mathbf{e}_1^* \cdot \mathbf{e}_{\mathcal{R}} = 0$, it follows that 
\begin{align}
\label{delta N formula}
\boldsymbol{\nabla}^\dag N & = \frac{1}{\sqrt{2\epsilon_*}} \left( \mathbf{e}_1^* + T_{\mathcal{R}\perp} \mathbf{e}_{\mathcal{R}} \right),
\end{align}
and therefore, the unit vector in the direction of $\boldsymbol{\nabla}^\dag N$ is
\begin{align}
\mathbf{e}_N = \cos \Delta_N \mathbf{e}_1^* + \sin \Delta_N \mathbf{e}_{\mathcal{R}}. 
\end{align}
These results generalize those found for two-field inflation in \cite{PetersonAndTegmark-2010b} and will be useful later ing calculating the non-Gaussianity arising from multifield inflation.

\subsection{Isocurvature and Cross Spectra}
\label{isocurv}

If there is more than one field present, there will also be a relic spectrum of isocurvatures modes and a cross spectrum between the density and isocurvatures modes.  Therefore, the detection of an isocurvature mode spectrum arising from inflation would indicate that at least two fields were present during inflation.   

Note, however, that unlike for the curvature modes, determining the isocurvature spectrum after inflation ends is more complicated because the isocurvature modes may decay further.  Such post-inflationary processing is highly model dependent and depends on the dynamics of reheating.  To make our discussion as broadly applicable as possible, we focus on the amplitudes of the isocurvature modes at the end of inflation, which can be construed as upper limits on the mode amplitudes.    Any post-inflationary model-dependent processing of the isocurvature modes is to be tacked onto these results by extending the transfer functions to encompass the whole evolution of the modes from horizon exit to the present era.  This can be represented by introducing a prefactor in the spectra in Eq. (\ref{isocurv spec}) and additional scale-dependent terms in the spectral indices for the isocurvature and cross spectra.     

Using some prior results from \cite{NibbelinkAndVanTent-2001,PetersonAndTegmark-2010}, the isocurvature spectrum at the end of inflation can be written as 
\begin{align}
\label{isocurv spec}
\mathcal{P}_{\mathcal{S}} = \left(\frac{H_*}{2\pi}\right)^2 \frac{1}{2\epsilon_*} \left\vert \mathbf{T}_{\mathcal{S}\perp} \right\vert^2,
\end{align}
where $\mathbf{T}_{\mathcal{S}\perp}$ is given by Eq. (\ref{multifield TRS v1}) and is calculated at the end of inflation.  

How the geometry and kinematics of inflation affects the isocurvature spectrum follows from our detailed discussion of the mode sourcing in Sections \ref{no sourcing}-\ref{sourcing} and \ref{transfer}.  So we focus on how the number of fields is reflected in the isocurvature spectrum.  Like for the other transfer function, we can break $\mathbf{T}_{\mathcal{S}\perp}$ into two parts:
\begin{align}
\label{e_S unit vec}
\mathbf{e}_{\mathcal{S}} \equiv \frac{\mathbf{T}_{\mathcal{S}\perp}}{|\mathbf{T}_{\mathcal{S}\perp}|}, \nonumber \\
T_{\mathcal{S}\perp} \equiv \left|\mathbf{T}_{\mathcal{S}\perp}\right|.
\end{align}
In the case of two-field inflation, $\mathbf{e}_{\mathcal{S}} = \mathbf{e}_2^*$, whereas for inflation with three or more effective fields, $\mathbf{e}_{\mathcal{S}} \neq \mathbf{e}_2^*$.  Using these two quantities, the multifield isocurvature spectrum becomes
\begin{align}
\label{iso spec}
\mathcal{P}_{\mathcal{S}} =  \left(\frac{H_*}{2\pi}\right)^2 \frac{1}{2\epsilon_*} T_{\mathcal{S}\perp}^2.
\end{align}
Like for the curvature spectrum, the expression for the multifield isocurvature spectrum has the same form as in the two-field case and therefore does not provide us any insight into the number of fields present during  multifield inflation, at least not to lowest-order in the slow-roll expansion.

Also if inflation is multifield, there will be a cross spectrum between the curvature and isocurvature modes, representing the mode correlations.  Combining results from \cite{NibbelinkAndVanTent-2001,PetersonAndTegmark-2010}, we can write the cross spectrum as
\begin{align}
\mathcal{C}_{\mathcal{RS}} & = \left(\frac{H_*}{2\pi}\right)^2 \frac{1}{2\epsilon_*} \left(\mathbf{T}_{\mathcal{R}\perp} \cdot \mathbf{T}_{\mathcal{S}\perp}\right).
\end{align}
Using Eqs. (\ref{e_R unit vec}) and (\ref{e_S unit vec}), this becomes
\begin{align}
\label{cross spec}
\mathcal{C}_{\mathcal{RS}}  & =  \left(\frac{H_*}{2\pi}\right)^2 \frac{1}{2\epsilon_*} T_{\mathcal{R}\perp} T_{\mathcal{S}\perp} \left(\mathbf{e}_{\mathcal{R}} \cdot \mathbf{e}_{\mathcal{S}} \right).
\end{align}
Comparing the above result to the two-field result in \cite{PetersonAndTegmark-2010}, we see that the results are identical with the exception of the term $\mathbf{e}_{\mathcal{R}} \cdot \mathbf{e}_{\mathcal{S}}$.  This is the first instance of a spectral quantity whose expression differs from the two-field case.

We can therefore use the cross spectrum to devise a test that will distinguish two-field inflation from inflation with three or more effective fields.  In analogy to the tensor-to-scalar ratio, the cross-correlation ratio \cite{BartoloEtAl-2001a} is
\begin{align}
r_C \equiv \frac{\mathcal{C}_{\mathcal{RS}}}{\sqrt{\mathcal{P}_{\mathcal{R}} P_{\mathcal{S}}}}.
\end{align}
Substituting Eqs. (\ref{two-field curv spec}), (\ref{iso spec}), and (\ref{cross spec}) yields
\begin{align}
\label{r_C result 1}
r_C = \sin \Delta_N \mathbf{e}_{\mathcal{R}} \cdot \mathbf{e}_{\mathcal{S}}.
\end{align}
If inflation is effectively two-field, then $\mathbf{e}_{\mathcal{R}} = \mathbf{e}_{\mathcal{S}} = \mathbf{e}_2$ and $r_C = \sin \Delta_N$.    But if $\mathbf{e}_{\mathcal{R}} \neq \mathbf{e}_{\mathcal{S}}$, then $r_C < \sin \Delta_N$,  signaling the presence of three or more effective fields.  

\begin{figure}[t]
\centering
\renewcommand{\arraystretch}{1.5}
\begin{tabular}{|x{1.5in}|x{1.5in}|} \hline
\multicolumn{2}{|c|}{\normalsize \rule{0cm}{0.5cm} Multifield Observables} \tabularnewline[0.5ex] \hline
\multicolumn{2}{|c|}{$\, $}\tabularnewline [-3.7ex] \hline 
\rule{0cm}{0.55cm} \normalsize $\beta_1 \equiv - \frac{r_T}{8n_T}$ & \normalsize $\beta_2 \equiv \frac{r_C}{\sqrt{1-\beta_1}}$ \tabularnewline [1.4ex] \hline
\multicolumn{2}{c}{$\, $}
\end{tabular}
\centerline{\includegraphics[width=97mm]{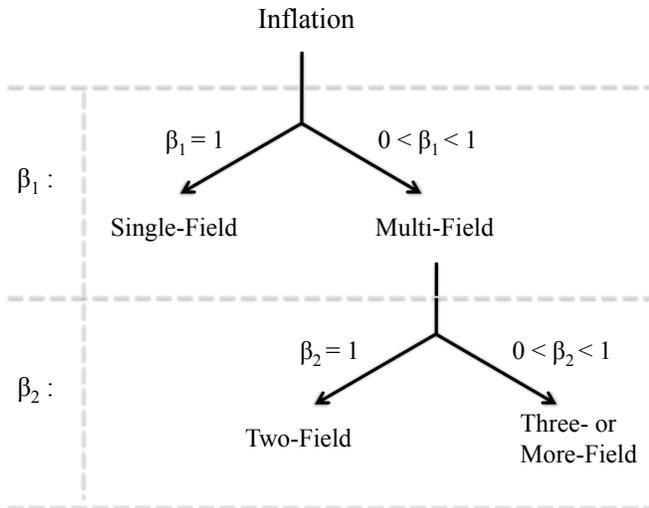}}
\caption{Multifield observables $\beta_1$ and $\beta_2$ indicate the effective number of fields during inflation.  \label{multifield Flow Chart}}
\end{figure}

Eq. (\ref{r_C result 1}) can also be cast solely in terms of spectral observables.   Substituting Eq. (\ref{rT inequality}) into Eq. (\ref{r_C result 1}) yields
\begin{align}
r_C \le \sqrt{1 + \frac{r_T}{8n_T}},
\end{align}
where the equality is satisfied when inflation can be described by two effective fields.  We can therefore define the following duo of \textit{multifield parameters}
\begin{eqnarray}
\label{betas}
\beta_1&\equiv&-{r_T\over 8 n_T}, \nonumber \\
\beta_2&\equiv&{r_C\over \sqrt{1 + \frac{r_T}{8n_T}}}.
\end{eqnarray}
The first multifield parameter, $\beta_1$, distinguishes multifield inflation from single-field inflation; it is derived from the well-known single-field consistency relation in Eq. (\ref{single-field rT consistency reln}). When $\beta_1=1$, inflation is single-field, whereas if $0 < \beta_1 < 1$, inflation is multifield.    The second multifield parameter, $\beta_2$, differentiates two-field models from models with three or more fields.  When $\beta_2=1$, inflation is driven by two effective fields, whereas for models with three or more effective fields, $ 0 < \beta_2 < 1$.   Moreover, these results remain valid even if the isocurvature modes decay after inflation, provided that the isocurvature and cross spectra are still detectable.\footnote{The one technical exception to the rule is if the decay of isocurvature modes takes $\mathbf{e}_{\mathcal{S}}$ from being not parallel to $\mathbf{e}_{\mathcal{R}}$ at the end of inflation to being parallel to $\mathbf{e}_{\mathcal{R}}$ at recombination, in which case there would appear to be only two effective fields, instead of at least three.  But this is a highly unlikely decay scenario.}  The reason why is because the result in Eq. (\ref{betas}) depends only on the structure of the transfer matrix formalism, not on the precise dynamics of the modes; these results apply in general to any scenario that can be described by the transfer matrix formalism.  This includes the curvaton model and inhomogeneous reheating, which both involve a very light field present during inflation that hugely sources and thus is said to generate the curvature perturbation following inflation.  However, if all of the isocurvature modes decay away completely or are undetectable---as in the case of complete thermalization after inflation---then both the isocurvature and cross-spectra will be unmeasurable and $\beta_2$ will be undefined.  In this case, the power spectra can only be used to distinguish single-field models from multifield models.  These results are summarized in Fig. \ref{multifield Flow Chart}. 

\noindent

\subsection{Higher-Order Spectra}
\label{nonGauss}

Finally, we consider whether higher-order spectra arising from Fourier transforms of higher-order mode correlation functions provide any clues about the number of fields present during inflation.  These higher-order spectra represent the non-Gaussian behavior  of the curvature perturbations.  The two lowest-order correlation functions are known as the bispectrum and trispectrum, respectively.   For standard multifield inflation, the local forms of these spectra predominate,\footnote{However, when $\mathbf{R} \neq \mathbf{0}$, other forms of the bispectra and trispectra may also be important.  We only give the local form here.}  with the local bispectrum represented by the parameter $f_{NL}$ and the trispectrum by the parameters $\tau_{NL}$ and $g_{NL}$.  For multifield inflation with canonical kinetic terms, the $\delta N$ formalism has been used to recast correlators of $\mathcal{R}$ in terms of partial derivatives of $N$ \cite{LythAndRodriguez-2005,AldaiaAndWands-2005,ByrnesEtAl-2006}.  We contend that with the substitution of covariant derivatives for partial derivatives, the same expressions apply in general multifield inflation with a curved field metric, giving 
\begin{align}
-\frac{6}{5} f_{NL}^{(4)} & = \frac{\mathbf{e}_N^\dag \, \boldsymbol{\nabla}^\dag \boldsymbol{\nabla} N \mathbf{e}_N}{|\boldsymbol{\nabla} N|^2}, \nonumber \\
\label{non-Gauss delta N formulas}
\tau_{NL} & = \frac{\mathbf{e}_N^\dag \, \boldsymbol{\nabla}^\dag \boldsymbol{\nabla} N \, \boldsymbol{\nabla}^\dag \boldsymbol{\nabla} N \,  \mathbf{e}_N}{|\boldsymbol{\nabla} N|^4} ,\\
\frac{54}{25} g_{NL} & = \frac{\mathbf{e}_N^\dag \, \boldsymbol{\nabla}^\dag \boldsymbol{\nabla}  \boldsymbol{\nabla}  N \, \mathbf{e}_N  \, \mathbf{e}_N}{|\boldsymbol{\nabla} N|^3}. \nonumber
\end{align}
Our contention has recently been confirmed and proved in much more detail by \cite{EllistonEtAl-2012}, which we refer the interested reader to.  

Consider first $f_{NL}$.  Our equation for $f_{NL}$ includes only the $k$-independent part, $f^{(4)}_{NL}$, which is the part of the local bispectrum that arises from the superhorizon evolution of nonlinearities \cite{VernizziAndWands-2006}; we ignore the undetectably small contribution from the $k$-dependent part, $f_{NL}^{(3)}$, which satisfies the bound $|-\frac{6}{5}f_{NL}^{(3)}| \le \frac{11}{96} r_T$  \cite{LythAndZaballa-2005,VernizziAndWands-2006}.   For $f_{NL}^{(4)}$, we calculated an expression for it in two-field inflation \cite{PetersonAndTegmark-2010b} using the spectral observables and by operating $\boldsymbol{\nabla}$ on the transfer function expression for $\boldsymbol{\nabla}^\dag N$ in Eq. (\ref{delta N formula}).  The calculation is similar for the case of general multifield inflation, so repeating the steps outlined in \cite{PetersonAndTegmark-2010b}, the bispectrum parameter can be written as:
\begin{widetext}
\begin{align}
\label{fNL general expression}
-\frac{6}{5}f_{NL}^{(4)} = \frac{1}{2} \cos^2 \Delta_N & (n_{\mathcal{R}} - n_T) + \sin \Delta_N \cos \Delta_N \left[(\mathbf{e}_{\mathcal{R}} ^\dag M \mathbf{e}_1)^* + \sin \Delta_N \cos \Delta_N \, \sqrt{-n_T} \, \mathbf{e}_{\mathcal{R}}  \cdot \boldsymbol{\nabla} T_{\mathcal{R}\perp}  \right].  
\end{align}
\end{widetext}
Eq. (\ref{fNL general expression}) is largely a formal equation, but nonetheless it can be used to determine whether the bispectrum parameter reveals the number of fields active during inflation.   In single-field inflation, $\mathbf{e}_{\mathcal{R}}$ vanishes because $T_{\mathcal{R\perp}} = 0$, yielding the single-field consistency relation $-\frac{6}{5}f_{NL}^{(4)} = \frac{1}{2} (n_{\mathcal{R}} - n_T)$,\cite{CreminelliAndZaldarriaga-2004}\footnote{The standard single-field consistency relation for $f_{NL}$ includes contributions from both $f_{NL}^{(3)}$ and $f_{NL}^{(4)}$.  When both contributions are included, the single-field result for the local bispectrum is  $-\frac{6}{5}f_{NL} = \frac{1}{2}  n_{\mathcal{R}}$  \cite{CreminelliAndZaldarriaga-2004}.} which is below the detection threshold.  In multifield inflation,  all terms except for $\mathbf{e}_{\mathcal{R}}  \cdot \boldsymbol{\nabla} T_{\mathcal{R}\perp}$  will be undetectably small, and the only difference between the above result and the result for two-field inflation is that $\mathbf{e}_2^*$ has been replaced by $\mathbf{e}_{\mathcal{R}}$.   So unless $\mathbf{T}_{\mathcal{R\perp}}$ is known, $f_{NL}$ cannot be used to distinguish two-field inflation from inflation with three or more fields.

As an aside, the formal expression in Eq. (\ref{fNL general expression}) can be used semianalytically if the transfer function $\boldsymbol{T}_{\mathcal{R}\perp}$ is computed in a small neighborhood about the field trajectory.   Also, it can be used to gain intuition into the expected magnitude of non-Gaussianity.  We demonstrated this for the case of two-field inflation in \cite{PetersonAndTegmark-2010b}.   For example, if the sourcing of curvature modes is small (i.e., $T_{\mathcal{R}\perp} \ll 1$), but $\boldsymbol{\nabla}  T_{\mathcal{R}\perp}$ varies dramatically in a direction orthogonal to the field trajectory, then from these equations, one can conclude that $f_{NL}$ will be large and that $\tau_{NL} \gg f_{NL}^2$.   Such a scenario arises when the field trajectory rolls along a ridge in the inflationary potential.  Eq. (\ref{fNL general expression}) is therefore useful because it tells us that similar conditions of instability in the inflationary trajectory are needed for large non-Gaussianity.  

Next, we find the trispectrum parameters.  First, in the single-field limit,  $\tau_{NL} =  \left(\frac{6}{5}f_{NL}^{(4)}\right)^2$ and hence is undetectably small.  This expression represents a consistency relation for single-field inflation \cite{SuyamaAndYamaguchi-2007}.  For the multifield case, following the steps outlined in \cite{PetersonAndTegmark-2010b}, we obtain
\begin{align}
\label{tau in terms of obs}
\tau_{NL} = & \frac{1}{\sin^2 \Delta_N} \left[\frac{6}{5}f_{NL}^{(4)} + \frac{1}{2} \cos^2 \Delta_N  (n_{\mathcal{R}}-n_T) \, \right]^2 \nonumber \\ & + \frac{1}{4} \cos^2 \Delta_N \, \left(n_{\mathcal{R}}-n_T\right)^2.
\end{align}
This expression for general multifield inflation is identical to the corresponding expression for two-field inflation.   Thus the trispectrum parameter $\tau_{NL}$ cannot distinguish two-field inflation from multifield inflation with more fields.   But $\tau_{NL}$ can be written completely in terms of other spectral observables.  Using Eq. (\ref{tau in terms of obs}) and that 
\begin{align}
\label{cos delta N in terms of spec obs}
-\frac{r_T}{8n_T} = \cos^2 \Delta_N,
\end{align}
$\tau_{NL}$ can be written as
\begin{align}
\label{new consist cond}
\tau_{NL} = & \frac{1}{1+\frac{r_T}{8n_T}} \left[\frac{6}{5}f_{NL}^{(4)} - \frac{1}{2} \frac{r_T}{8n_T}  (n_{\mathcal{R}}-n_T) \, \right]^2  \nonumber \\ & - \frac{1}{4} \frac{r_T}{8n_T}\, \left(n_{\mathcal{R}}-n_T\right)^2,
\end{align}
which we note is only valid when inflation contains multiple fields.  Eq. (\ref{new consist cond}) represents a new consistency condition for general multifield inflation.  In the limit where $f_{NL}$ is detectably large (i.e., $|f_{NL}| \gtrsim 3$), the above multifield consistency condition reduces to
\begin{align}
\label{tNLeq}
\tau_{NL}  = \frac{1}{1+\frac{r_T}{8n_T}} \left(\frac{6}{5} f_{NL}\right)^2.
\end{align}
In this limit, the value of $\tau_{NL}$ relative to $f_{NL}^2$ is controlled solely by the ratio of $r_T$ to $n_T$;  the larger the sourcing of the curvature modes by the isocurvature modes, the more $\tau_{NL}$ approaches $\left(\frac{6}{5} f_{NL}\right)^2$.  In other words, only multifield inflationary scenarios where the multifield effects are very weak can produce $\tau_{NL} \gg f_{NL}^2$.  This observation and Eqs. (\ref{new consist cond}) and (\ref{tNLeq}) represent new findings applicable to general multifield inflation.  And the size of  $\tau_{NL}$ relative to $f_{NL}^2$ in Eq. (\ref{tNLeq}) follows from the kinematics of the background trajectory and an analysis of the effective mass matrix over the trajectory, again reflecting how the geometry of the inflationary Lagrangian affects the spectra.

Lastly, for the trispectrum parameter $g_{NL}$, we follow the steps in \cite{PetersonAndTegmark-2010b} to obtain:
\begin{align}
\frac{54}{25}  g_{NL} = & - 2 \tau_{NL} + 4 \left(\frac{6}{5} f_{NL}^{(4)}\right)^2 \nonumber \\ & + \sqrt{\frac{r_T}{8}} \, \mathbf{e}_N \cdot \boldsymbol{\nabla} \left(-\frac{6}{5}f_{NL}^{(4)}\right).
\end{align}
As written, the above result for multifield inflation is very formal, but since it is identical in form to that in two-field inflation, it tells us that $g_{NL}$ can only be used to distinguish single-field inflation from multifield inflation.   In the case of single-field inflation, the above expression reduces to
\begin{align}
\frac{54}{25}  g_{NL} & = 2 \left(\frac{6}{5} f_{NL}^{(4)}\right)^2 + \left(-\frac{6}{5}f_{NL}^{(4)}\right)', 
\end{align}
where $\frac{d f_{NL}}{d \ln k} \approx f_{NL}'$ represents the scale dependence of $f_{NL}$ and where we used the single-field limit of Eq. (\ref{tau in terms of obs}) to obtain the last relation.  

In sum, detection of non-Gaussianity arising from the curvature modes would indicate that inflation is multifield, but cannot otherwise provide insight into the effective number of fields present during inflation.  The reason why is because the multifield inflation expressions for the non-Gaussian parameters are identical to those in two-field inflation after the replacement $\mathbf{e}_2^* \rightarrow \mathbf{e}_{\mathcal{R}}$, and hence they cannot differentiate models with two fields from those with three or more fields. But fortunately, combining observables from the tensor, curvature, isocurvature, and cross spectra can in principle be used to distinguish among inflationary models driven by one, two, and three or more fields, as summarized in Fig. \ref{multifield Flow Chart}.

\section{Conclusions}

The interactions among the field perturbations in multifield inflation are determined by the geometric properties of the inflationary potential and the field manifold.   Because the mode interactions serve as the critical bridge between the inflationary Lagrangian and the cosmic observables, they can be used to compare inflationary models based on common geometric features that cut across several types of Lagrangians.  For example, Lagrangians that give rise to a field trajectory that turns sharply in field space tend to have highly scale-dependent curvature spectra \cite{PetersonAndTegmark-2010}, while those that produce a field trajectory that rolls along a ridge in the potential are more likely to produce large non-Gaussianity, all else being equal \cite{PetersonAndTegmark-2010b}.    

It is therefore critical to develop tools to understand how the mode interactions reflect the geometric properties of the inflationary Lagrangian.  While the mode interactions are well understood in the case of general two-field inflation and in some cases of multifield potentials, they  are not well understood for an arbitrary multifield Lagrangian.   Instead the $\delta N$ formalism has been heavily relied on to calculate the spectra, which although powerful,  does not provide much insight into the evolution of modes.   In this manuscript, we attempted to extend previous work to uncover how the geometric and kinematical features of the Lagrangian affect the interactions among  modes, how this determines the effective number of active fields during inflation, and how this is reflected in the spectral observables.   

We started in Section \ref{Background Fields} by presenting the covariant equation of motion for the fields and by delineating a framework to parse the field vector kinematics.  The kinematics of the background fields induce a basis called the kinematical basis and a matrix of turn rates, $\mathbf{Z}$, which characterizes how quickly these basis vectors are rotating.  We concluded our treatment of the background fields by discussing underappreciated subtleties of the slow-roll limit when multiple scalar fields are present.

In Section \ref{Perbs}, we explored the equations of motion for the field perturbations in both the given and kinematical bases and showed how the evolution of modes reflects the geometry of the Lagrangian.  In the combined superhorizon and slow-roll limits, the equation of motion for the field perturbations depends only on the effective mass matrix $\mathbf{\tilde{M}}$---which represents the covariant Hessian of the potential and the Riemann tensor of the field manifold---and the turn rate matrix $\mathbf{Z}$.   We then studied the mode interactions one by one in the kinematical basis.   We started by considering the evolution of the $\delta \phi_n$ mode in the absence of sourcing, and we discussed how the concavity of the potential and the curvature of the field manifold determine that mode's instrinic evolution.   In analogy to the adiabatic conservation law in single-field inflation, we showed that there are up to $d$ mode-related quantities in $d$-field inflation that may be conserved.  

Next, we looked at sourcing.  For quadratic potentials with canonical kinetic terms, the mode equations simplify radically, in a way such that each mode $\delta \phi_n$ can be sourced only by $\delta \phi_{n+1}$ but only when the basis vector $\mathbf{e}_n$ is turning into the direction of $\mathbf{e}_{n+1}$.  For this special class of models, all turn rate matrix coefficients can be expressed in terms of the mass matrix, and all mode sourcing equations assume the same form as for the adiabatic mode.   We then used this special case as a reference point for the discussion of mode sourcing in the case of an arbitrary Lagrangian.  We argued that the mode interactions in a general inflation model can be divided into features shared in common with canonical quadratic models and features that arise from higher-order covariant derivations of the potential and corrections from the field metric, and we advocated this approach as way to gain greater insight into how differences in Lagrangians translate into differences in the cosmic observables.   In parallel, we discussed the three types of sourcing terms: two are geometrical terms and one is kinematical.  The geometrical terms involve off-diagonal terms in both the covariant Hessian of the potential and in the Riemann tensor of the field metric contracted with two instances of $\mathbf{e}_1$ and modulated by $\epsilon$, and we interpreted these terms geometrically.  The kinematical terms are simply the turn rates of $\mathbf{e}_n$ into the $\mathbf{e}_{n+1}$ and $\mathbf{e}_{n-1}$ directions and can intuitively understand as gains and losses in the amplitude of $\delta \phi_n$ due to the rotation of basis.  We also gave several examples of how inferences about the mode sourcing can be made by determining the geometric and kinematical features of a Lagrangian.
  
With this in mind, we focused in Section \ref{CosmicObs} on how the Lagrangian geometry and kinematics determines the effective number of fields and how this number is reflected in the power spectra, bispectrum, and trispectrum.  We pointed out that the effective number of fields needed to describe the background and perturbed solutions do not necessarily coincide, and we gave a method to determine the effective dimension of a multifield system in the slow-roll limit.  Next, we presented known formulas for the power spectra and generalized the two-field expressions for the local non-Gaussianity parameters to multifield inflation.  We found a new multifield consistency relation among $\tau_{NL}$, $f_{NL}$, $r_T$, and $n_T$ for detectably large non-Gaussianity to multifield inflation, and we discovered a \textit{multifield observable} involving the cross spectrum that can potentially distinguish two-field models from models with three or more effective fields.    This result expressed is independent of post-inflationary processing of the modes.  However, the caveat is that all four spectra must be detectably large and hence they do not apply in the case of scenarios such as complete thermalization after inflation. 

Stepping back and looking at the big picture, since more sensitive measurements of the spectral observables, along with new spectral observables, will reveal further clues into the nature of inflation, we must be posed to extract phenomenological information from these measurements.   Since it is impractical to test the myriad inflationary scenarios one by one against these measurements, it is important that we study types of geometric and kinematical features that arise from inflationary Lagrangians and determine how these features affect the cosmic observables.  This will allow us to work backwards from constraints on the cosmic observables to identify the key features of the inflationary Lagrangian that described our early Universe.   The work presented in this paper represents a step forward towards this goal.

\section{Acknowledgments}

This work was supported by an NSF Graduate Research Fellowship, NSF grants AST-0708534 and AST-0908848, and a fellowship from the David and Lucile Packard Foundation.

\end{document}